\documentclass[12pt]{iopart}
\newcommand{\gguide}{{\it Preparing graphics for IOP journals}}
\usepackage{latexsym}
\usepackage{graphics}

\usepackage{iopams}
\usepackage{latexsym}
\usepackage{graphics}
\usepackage{float}

\usepackage{graphicx}
\DeclareGraphicsExtensions{.pdf,.png,.jpg}

\usepackage{iopams}

\begin{document}

\title[Cospectral quantum graphs]
{Cospectral quantum graphs}

\author[A. Chernyshenko]{A. Chernyshenko}

\address{South Ukrainian National Pedagogical University, Staroportofrankovskaya str. 26, Odesa, Ukraine, 65020
}

\ead{nastya.chernyshenko12@gmail.com}

\author{V. Pivovarchik}


\address{South Ukrainian National Pedagogical University,
Staroportofrankovskaya str., 26, Odesa, Ukraine, 65020}
\ead{vpivovarchik@gmail.com}

\begin{abstract}
Spectral problems are considered generated by the Sturm-Liouville equation on connected simple equilateral graphs with the Neumann and  Dirichlet boundary conditions at the   pendant vertices and continuity and Kirchhoff's conditions at the interior vertices.  We highlight the cases where the first and the second terms of the asymptotics of the eigenvalues uniquely determine the shape of the graph or of its interior subgraph.

\end{abstract}

\vspace{2pc} \noindent{\it Keywords}: Dirichlet boundary
condition, Neumann boundary condition,  Kirchhoff's condition, spectrum, tree.

\submitto{Journal of Physics A}
\maketitle

\section{Introduction}

Cospectrality is an old problem of classical graph theory. In \cite{CS} an example of cospectral graphs was given (see Fig. 1).

\begin{figure}[H]
  \begin{center}
    \includegraphics[scale= 0.88 ] {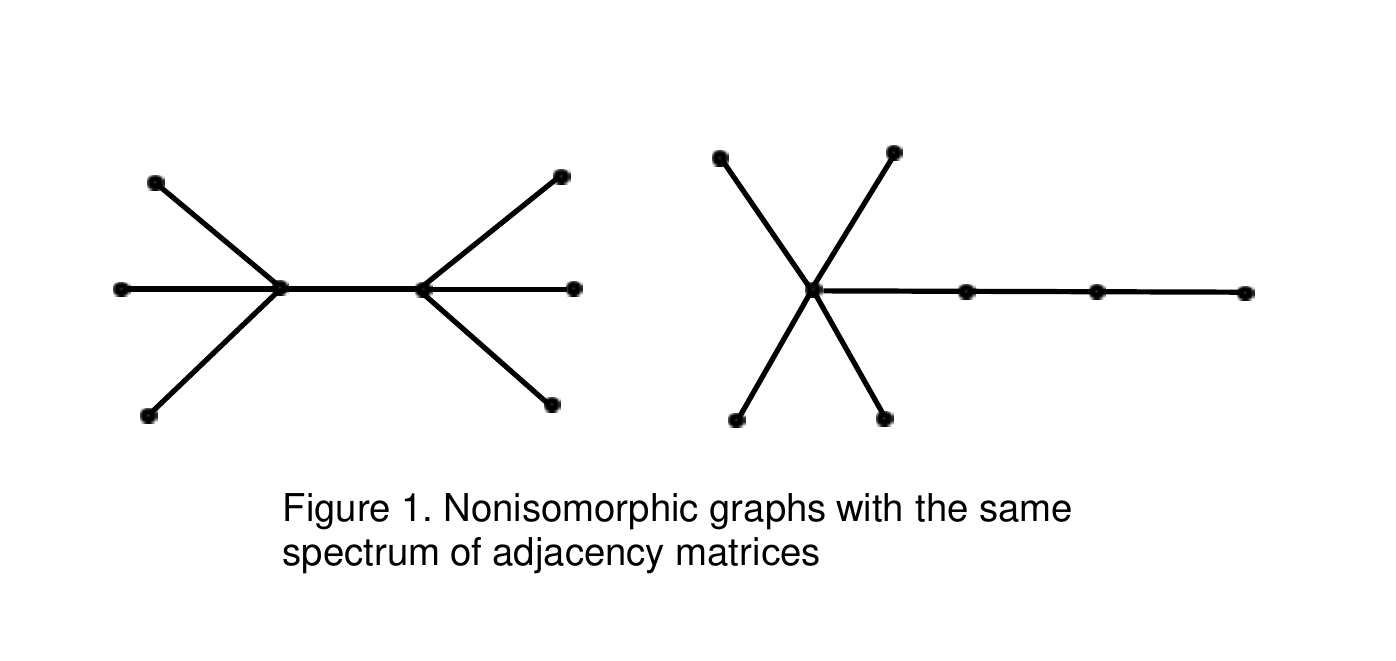}
   \end{center}
\end{figure}

 By cospectral (isospectral) in classical graph theory they mean two nonisomorphic graphs with the same spectrum of the adjacency matrix (see \cite{CDS}, Sec. 6.1).  In many cases it is more convenient to deal with  discrete Laplacians instead of adjacency matrices. There are different definitions of the discrete Laplacian sometimes called normalized Laplacian (see \cite{FanChung}, p.2).  By discrete Laplacian we mean the matrix  $D^{-1/2}AD^{-1/2}$ where $A$ is the adjacency matrix and $D=diag\{d(v_1), d(v_2),...,d(v_p)\}$ where $d(v_j)$ is the degree of the vertex $v_j$. This definition differs from the definition in \cite{FanChung} by shifting the spectral parameter $\lambda\rightarrow \lambda +1$. Despite the spectra of the adjacency matrices  of the graphs shown at Fig. 1 are the same, 
the spectra of the discrete Laplacians of these graphs are different: they are the sets of zeros of the polynomials
$
16z^8-25z^6+9z^4
$
and
$
20z^8-33z^6+13z^4,
$
respectively.

In quantum graph theory it is usual to   consider  spectral problems generated by the Sturm-Liouville equations on  equilateral metric  graph domains with the
Neumann or Dirichlet boundary conditions at the graphs  pendant vertices and generalized Neumann (continuity and Kirchhoff's) conditions at its interior vertices. Here the problem of cospectrality arises also.

 It was shown in \cite{vB} that there exist cospectral  graphs (nonisomorphic graphs with the same spectrum of the Sturm-Liouville problem) in quantum graph theory.  The  example mentioned in \cite{vB} shows two nonisomorphic equilateral metric graphs 
of Fig. 2.   

\begin{figure}[H]
  \begin{center}
    \includegraphics[scale= 0.88 ] {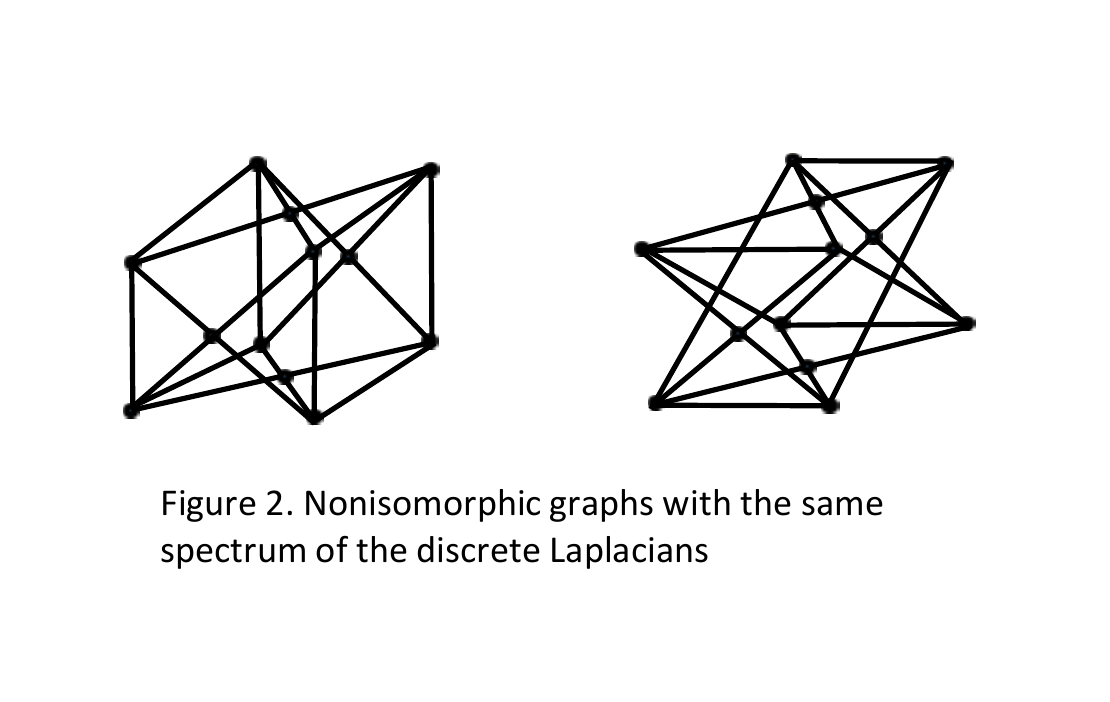}
   \end{center}
\end{figure}

It should be noticed that in the case of graphs with noncommensurate edges the spectrum uniquely determines the shape of the graph \cite{GS}. 
Notice that the graphs of Fig. 2 are regular and have the same number of edges. It is easy to see that two nonisomorphic regular graphs with zero potentials on the edges, the same number of the vertices and the same degree of the vertices have the same spectrum of the discrete Laplacian if and only if they have the same spectrum of their adjacency matrices. But this is true only for regular graphs. 

The spectra of quantum graph problems are related to the discrete Laplacians of the corresponding graphs in the following way: the eigenvalues of the discrete Laplacian are nothing but the coefficients of the second term of the asymptotics of the eigenvalues of  the quantum problem with the (generalized) Neumann boundary conditions at the vertices of this graph (see \cite{CP} where the results of \cite{Ca}, \cite{Ex} and of \cite{CaP} were used). This enables to obtain information on the form of a graph using the asymptotics of the eigenvalues.  
 
In \cite{KN}, \cite{BKS} a `geometric' Ambarzumian's theorem was proved stating that if the spectrum of the Sturm-Liouville problem with  the Neumann boundary conditions is such as in the case of the problem on a single interval with the zero   potential  then the graph is $P_2$ and  the potential of the Sturm-Liouville equation is zero. In \cite{CP} a geometric Ambarzumian's theorem was proved for connected simple equilateral graphs of 5 or less vertices and for trees of 8 or less vertices. This theorem states that if the spectrum of the Sturm-Liouville problem with the Neumann boundary conditions at the pendant vertices and generalized Neumann conditions at the interior vertices is such as the spectrum of this problem in case of zero potentials on the edges then  this spectrum uniquely determines the form of the graph and the zero potentials on the edges. 

However, this result cannot be extended to the case of connected simple equilateral graphs of 6 vertices. The graphs shown at Fig. 3 have the same first and second terms of asymptotics of the eigenvalues (see Theorem 4.5 below). These graphs were obtained in \cite{KM} by chopping a vertex of $K_5$ into two vertices.

\begin{figure}[H]
  \begin{center}
    \includegraphics[scale= 0.88 ] {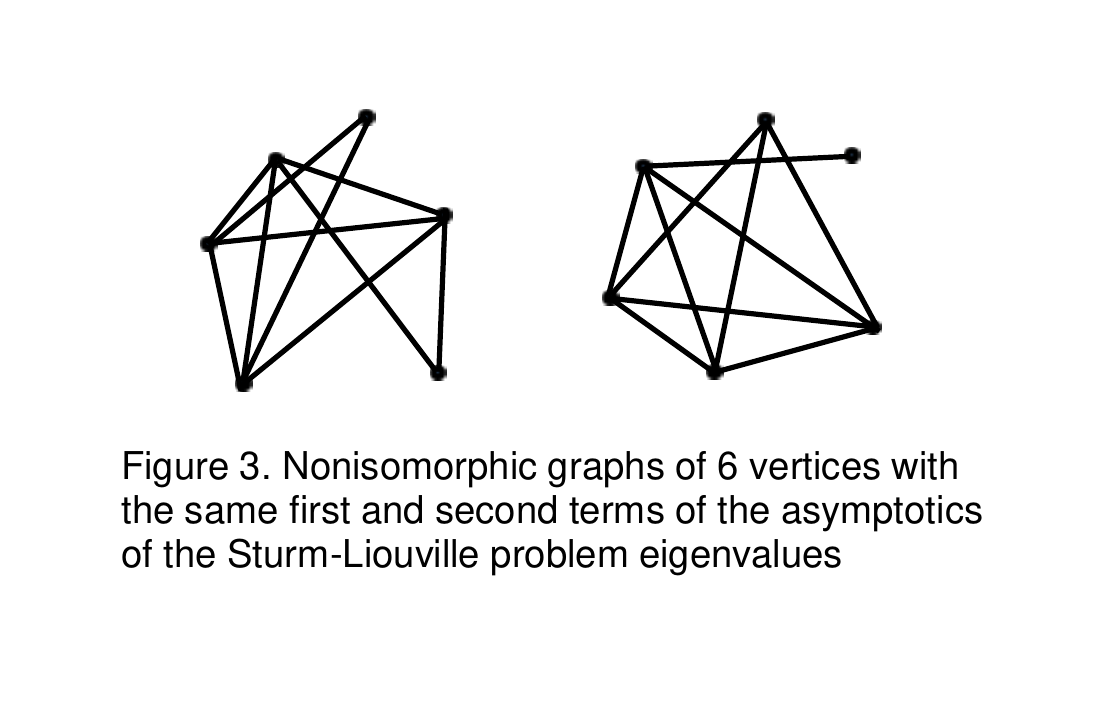}
   \end{center}
\end{figure}

It appears that the identity of the discrete Laplacians spectra does not mean identity of the first and the second terms of the asymptotics of the eigenvalues of the Sturm-Liouville problem. In addition we need the numbers of the edges of the graphs to be equal. For example the graphs shown at Fig. 4 have the same spectra of the discrete Laplacians but different the first and the second terms of the asymptotics of the eigenvalues of the Sturm-Liouville problem because their numbers of edges are different. Therefore, in what follows looking for cospectral graphs, we are interested in simple connected metric equilateral graphs with the same number edges and the same spectrum of  their discrete  Laplacians.

\begin{figure}[H]
 \begin{center}
    \includegraphics[scale= 0.88 ] {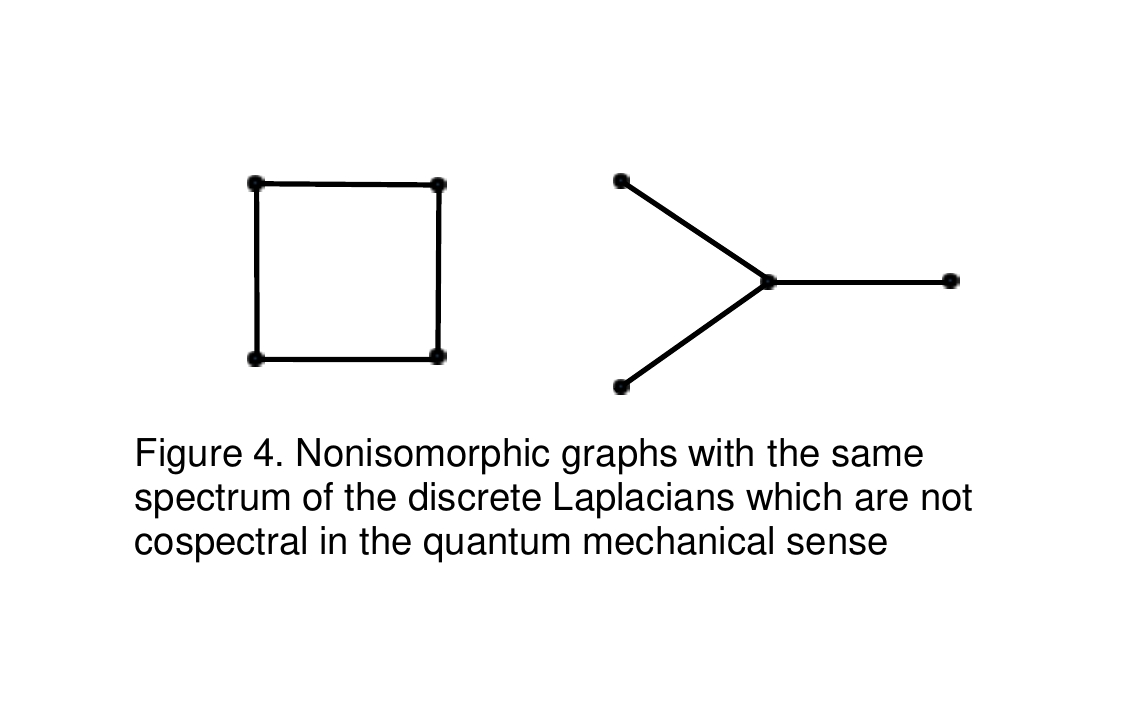}
   \end{center}
\end{figure}

It is known \cite{CP} that  the eigenvalues of the discrete Laplacian can be found from the asymptotics of eigenvalues of the  Sturm-Liouville problem on a graph  not only in the case of `Ambarzumian's' asymptotics. Thus, putting aside Ambarzumian's theorem and the potentials admitting them to  be arbitrary real $L_2$ functions we put a question: 
can we find the shape of a simple connected equilateral graph using the asymptotics of the eigenvalues of the Sturm-Liouville  spectral problem with the Neumann boundary conditions at the pendant vertices and generalized Neumann conditions at the interior vertices?   Also we are interested in what information on
 the shape of the graph can be obtained from the eigenvalue asymptotics in case of the Dirichlet conditions at the pendant vertices. In \cite{B}, \cite{B2} admitting Dirichlet conditions at some of the vertices a method for constructing families of cospectral systems is proposed, using linear representations of finite groups. 



  

 In Section 2 we formulate the spectral  Sturm-Liouville problem on a simple connected equilateral graph with the generalized Neumann conditions at the interior vertices and Neumann conditions at the pendant vertices and another problem with the Dirichlet conditions at the pendant vertices.

In Section 3 we give some auxiliary results. We consider an associated finite-dimensional spectral problem  and show the relation between the spectrum of the Sturm-Liouville problem on the graph and the spectrum of the discrete Laplacian corresponding to this graph in case of the Neumann conditions at the pendant vertices. Also we show the relations between the spectrum of the Sturm-Liouville problem and the spectrum of the modified discrete Laplacian of the graphs  interior subgraph in the case of the Dirichlet conditions at the pendant vertices.

In Section 4 we show that for simple connected equilateral graphs with the number of vertices $p\leq 5$ the first and the second terms of the asymptotics of the eigenvalues of the Sturm-Liouville problem with the Neumann boundary conditions at the pendant vertices and the generalized Neumann conditions at the interior vertices uniquely determine the shape of the graph. The same result appears to be true for trees with 8 or less vertices. We also show that among the graphs of 6 vertices there are two of the same the first and the second terms of asymptotics.

In Section 5 we consider the Sturm-Liouville spectral problem on a simple connected equilateral graph with the Dirichlet boundary conditions at the pendant vertices and generalized Neumann conditions at the interior vertices. Here we can receive information from the asymptotics sufficient to determine the shape of the graph only in the simpliest cases. However, we obtain much information on the shape of the interior subgraph, i. e. the subgraph obtained by deleting the pendant vertices and the edges incident with them.


\section{Statement of the problems}

Let $G$ be a simple connected  equilateral graph with $p$ vertices and $g$ edges of the length $l$ each. We denote by $v_i$ the vertices, by $d(v_i)$ their
degrees, by  $e_j$ the edges.  
We direct each peripheral (incident with the a pendant vertex) edge away from its pendant vertex. Orientation of the rest of the edges is arbitrary.
Thus,  for an interior vertex $v_i$ we consider its indegree by $d^+(v_i)$
and its outdegree  $d^-(v_i)=d(v_i)- d^+(v_i)$. Denote by $W^-(v_i)$ the set of indices $j_s$ ($s=1,2,...,d^-(v_i)$) of the edges outgoing from $v_i$ and by $W^+(v_i)$ the set of indices $k_s$ ($s=1,2,...,d^+(v_i)$) of the edges incoming into $v_i$.

 Local
coordinates for the edges identify each  edge $e_j$ with the interval
$[0,l]$ so that the local coordinate increases in the direction of the edge. This means that each pendant vertex has the
local coordinate $0$.

Each
interior vertex  has the local coordinate  $l$ on its incoming edge, while the local coordinate of the vertex is $0$ on each outgoing edge. Functions $y_j$ on the edges are
subject to the system of $g$ scalar Sturm-Liouville equations
\begin{equation}
-y_j^{\prime\prime}+q_j(x)y_j=\lambda y_j, \ \ (j=1, 2, ..., g)
\label{2.1}
\end{equation}
where $q_j$ is a real-valued function which belongs to
$L_2(0,l)$.  
For each interior vertex  with outgoing edges $e_j$ ($j\in W^-(v_i)$) and
incoming edges $e_k$ ($k\in W^+(v_i)$) the continuity conditions are
\begin{equation}
y_j(0)=y_k(l),  \label{2.2}
\end{equation}
 and  Kirchhoff's condition is
 \begin{equation}
\mathop{\sum}\limits_{k\in W^+(v_i)} y_k^{\prime}(l)=\mathop{\sum}\limits_{j\in W^-(v_i)} y_j^{\prime}(0).
 \label{2.3}
 \end{equation}
We impose the Neumann boundary conditions
\begin{equation}
 y^{\prime}_j(0)=0, \label{2.4}
\end{equation} 
at  the  pendant vertices $v_j$ ($j=r+1, r+2, ..., p_{pen}$) and
the Dirichlet  boundary conditions
\begin{equation}
\label{2.5}
y_j(0)=0
\end{equation}
at the rest of the pendant vertices $v_j$ ($j=1,2, ..., r$).

Let us denote by $s_j(\sqrt{\lambda},x)$ the solution of the
Sturm-Liouville equation (\ref{2.1}) on the edge $e_j$ which
satisfies the conditions
$s_j(\sqrt{\lambda},0)=s_j^{\prime}(\sqrt{\lambda},0)-1=0$ and by
$c_j(\sqrt{\lambda},x)$ the solution   which satisfies the conditions
$c_j(\sqrt{\lambda},0)-1=c_j^{\prime}(\sqrt{\lambda},0)=0$. Then the {\it
characteristic function} $\Phi(\lambda)$, i.e. an entire function whose set of zeros
coincides with the spectrum of the problem (\ref{2.1})-(\ref{2.5})   
can be expressed via
$s_j(\sqrt{\lambda},l)$, $s_j^{\prime}(\sqrt{\lambda},l)$,
$c_j(\sqrt{\lambda},l)$ and $c_j^{\prime}(\sqrt{\lambda},l)$. To do it we
introduce the following system of vector-functions
$\psi_j(\lambda,x)={\rm col}\{0,0,...,s_j(\sqrt{\lambda},x),...,0\}$ and
$\psi_{j+g}(\lambda,x)={\rm col}\{0,0,...,c_j(\sqrt{\lambda},x),...,0\}$
for $j=1,2,...,g$. As in
\cite{Pokor} we denote by $L_j$ ($j=1,2,...,2g$) the linear
functionals generated by (\ref{2.1})--(\ref{2.5}). Then
$\Phi(\lambda)=||L_j(\psi_k(\lambda,x)||_{j,k}^{2g}$ is the
characteristic matrix which represents the system of linear
equations describing the continuity and Kirchhoff's conditions for
the interior vertices. Then we call
 $$
 \phi(\lambda):=\det(\Phi(\lambda))
 $$
 the {\it characteristic
function}  of problem (\ref{2.1})--(\ref{2.5}). The characteristic
function is determined up to a constant multiple.

\section{Auxiliary results}
\setcounter{equation}{0} \hskip0.25in

 
For a simple graph, the matrix
$A=(A_{i,j})_{i,j=1}^p$ where $A_{i,i}=0$ for all $i=1,2,...,p$ and
for $i\not=j$:
\[\quad A_{i,j}=\left\{\begin{array}{c} 1 \  {\rm if } \ v_i \ {\rm and } \ v_j \ {\rm are  \ adjacent,} \\  0 \ {\rm otherwise}, \end{array}\right.\] 
is called the \textit{adjacency} matrix\index{adjacency matrix}.
Let
$$
D=diag\{d(v_1), d(v_2), ..., d(v_p)\}
$$
be the degree matrix. Then 
$$
\widetilde{ A}=D^{-1/2}AD^{-1/2}
$$
is called the \textit{weighted adjacency} matrix or \textit{discrete Laplacian}.

Let $G$ be a simple connected equilateral graph with $g\geq 2$ edges, $p$ vertices, $p_{pen}$ pendant vertices. Let $r$ ($0\leq r \leq p_{pen}$) be the number of pendant vertices with the Dirichlet conditions. We impose the Neumann conditions at the rest of the pendant vertices.  The graph $\hat{G}$ is obtained by removing the pendant vertices with Dirichlet boundary conditions and the edges  incident with them in $G$. For convenience, the vertices of $\hat{G}$ we denote by $v_{r+1}$, $v_{r+2}$,..., $v_{p}$. Let $\hat{A}$  be the adjacency matrix of ${\hat G}$, let
$\hat{D_G}=diag \{d(v_{r+1}), d(v_{r+2}), ..., d(v_{p})\}$, where  $d(v_i)$ is the degree of the vertex $v_i$ in $G$. We consider the polynomial $P_{G,\hat{G}}$ defined by 
$$
\label{3.2}
P_{G,\hat{G}}(z):=det(z\hat{D}_G-\hat{A}).
$$
The following theorem was proved in \cite{MP2} (Theorem 6.4.2).

{\bf Theorem 3.1} {\it Let $G$ be a simple connected graph with at least two edges. Assume that all edges have the same length $l$ and the same potentials symmetric with respect to the midpoints of the edges ($q(l-x)=q(x)$). Then the spectrum of problem (\ref{2.1})--(\ref{2.5})  coincides with the set of zeros of  the characteristic function
$$
\phi_D(\lambda)=s^{g-p+r}(\sqrt{\lambda},l)P_{G,\hat{G}}(c(\sqrt{\lambda},l))
$$
where $s(\sqrt{\lambda},x)$ and $c(\sqrt{\lambda},x)$ are the solutions of the Sturm-Liouville equation which satisfies $s(\sqrt{\lambda}, 0)=s'(\sqrt{\lambda}, 0)-1=0$ and $c(\sqrt{\lambda},0)-1=c'(\sqrt{\lambda},0)=0$. } 



\section{Spectral problem for the case of the Neumann boundary conditions at the pendant vertices}
\setcounter{equation}{0} \hskip0.25in

In this section we consider the case of Neumann boundary conditions at all the pendant vertices ($r=0$), i.e. we consider problem (\ref{2.1})--(\ref{2.4}).




{\bf Theorem 4.1} (Theorem 4.1 of \cite{CP}). {\it Let $g\geq p$. The eigenvalues of problem (\ref{2.1})--(\ref{2.3}), (\ref{2.5}) can be presented as the union of subsequences $\{\lambda_k\}_{k=1}^{\infty}=\mathop{\cup}\limits_{i=1}^{p+g}\{\lambda_k^{(i)}\}_{k=1}^{\infty}$
with the following asymptotics ($k\in\mathbb{N}:=\{1, 2, ... \}$).
\begin{equation}
\label{4.1}
\sqrt{\lambda_k^{(i)}}\mathop{=}\limits_{k\to\infty}\frac{2\pi (k-1)}{l}+\frac{1}{l}\arccos \alpha_{i}+O\left(\frac{1}{k}\right) \ \ {\rm for}  \ \  i=1,2,..., p,
\end{equation}
\begin{equation}
\label{4.2}
\sqrt{\lambda_k^{(i+p)}}\mathop{=}\limits_{k\to\infty}\frac{2\pi k}{l}
-
\frac{1}{l}\arccos \alpha_{i}+O\left(\frac{1}{k}\right) \ \  {\rm for} \ \  i=1, 2, ..., p,
\end{equation}
and if $q>p$
\begin{equation}
\label{4.3}
\sqrt{\lambda_k^{(2p+i)}}\mathop{=}\limits_{k\to\infty}\frac{\pi k}{l}+O\left(\frac{1}{k}\right) \ \  {\rm for} \ \ i=1, 2, ..., q-p \ \
\end{equation}
where $\{\alpha_i\}_{i=1}^p$ are the eigenvalues of the corresponding discrete Laplacian. }

Now let us see what information on the shape of a graph can be obtained from the asymptotics of the eigenvalues.  It is known that
\begin{equation}
\label{4.4}
\mathop{\lim}\limits_{k \to \infty}\frac{\lambda_k}{k^2}=\frac{\pi^2}{g^2l^2}
\end{equation}
(see \cite{Meh}, \cite{vB2}, \cite{Ni} or \cite{vB} Corollary 1). Thus, we can find the product $gl$.

{\bf Theorem 4.2} {\it Let the spectrum  $\{\lambda_k\}_{k=1}^{\infty}$ of problem (\ref{2.1})--(\ref{2.4}) for a simple connected equilateral graph with not more than 5 vertices consist of subsequences:
$\{\lambda_k\}_{k=1}^{\infty}=\mathop{\cup}\limits_{i=1}^{p+g}\{\lambda_k^{(i)}\}_{ k=1}^{\infty}$ which satisfy the asymptotics  
$$
\sqrt{\lambda_k^{(i)}}=\frac{2\pi (k-1)}{l}+\gamma_{i} +O\left(\frac{1}{k}\right) \ \ {\rm for}  \ \  i=1,2,..., p,
$$
$$
\sqrt{\lambda_k^{(i+p)}}=\frac{2\pi k}{l}
-
\gamma_{i} +O\left(\frac{1}{k}\right) \ \  {\rm for} \ \  i=1, 2, ..., p,
$$
and if $q>p$
$$
\sqrt{\lambda_k^{(2p+i)}}=\frac{\pi k}{l} +O\left(\frac{1}{k}\right)\ \  {\rm for} \ \ i=1, 2, ..., q-p \ \ 
$$
where $0<p\leq 5$, $p\leq q$, $\gamma\in\mathbb{R}$. 

Then the spectrum uniquely determines the shape of the graph and the length of an edge of this graph.} 

{\bf Proof} 
By Theorem 5.4 in \cite{CaP} we conclude that $|\lambda^{(j)}_k-{\tilde \lambda}^{(j)}_k|\leq C<\infty$ where  ${\tilde \lambda}^{(j)}_k$ are the eigenvalues of problem (\ref{2.1})--(\ref{2.4}) on the same graph with $q_j\equiv 0$ for all $j$ and therefore, presence of the potentials does not influence the first and the second terms of the asymptotics. By conditions of our theorem $\{\lambda_k\}_{k=1}^{\infty}=\mathop{\cup}\limits_{i=1}^{p+g}\{\lambda_k^{(i)}\}_{k=1}^{\infty}$ is the spectrum of problem (\ref{2.1})--(\ref{2.4}) on a simple connected graph, therefore the spectrum of the same problem with 0 potentials can be given as the union of subsequences with the asymptotics 
$$
\sqrt{\tilde{\lambda}_k^{(i)}}=\frac{2\pi (k-1)}{l}+\gamma_{i}  \ \ {\rm for}  \ \  i=1,2,..., p,
$$
$$
\sqrt{\tilde{\lambda}_k^{(i+p)}}=\frac{2\pi k}{l}
-
\gamma_{i} \ \  {\rm for} \ \  i=1, 2, ..., p,
$$
and if $q>p$
$$
\sqrt{\tilde{\lambda}_k^{(2p+i)}}=\frac{\pi k}{l} \ \  {\rm for} \ \ i=1, 2, ..., q-p \ \ 
$$
where $0<p\leq 5$, $p\leq q$. 
All possible spectra corresponding to simple connected graphs of 5 or less vertices consist of subsequences (\ref{4.1})--(\ref{4.3}), i.e. $\gamma_i=\frac{1}{l}arccos \ \alpha_i$
where $\{\alpha_i\}_{i=1}^p$ are the eigenvalues of the corresponding discrete Laplacians, i.e. the zeros of one of the polynomials mentioned in \cite{CP}. The list of the corresponding graphs is also given in \cite{CP}.

Thus, the set $\{\alpha_i\}_{i=1}^p$ uniquely determines the shape of the graph if $p\leq 5$. 
 and  in this case also the set $\{\alpha_i\}_{i=1}^p$ together with the number of edges $g$ can be found from the asymptotics. Then the length $l$ of an edge can be found from (\ref{4.4}).
These data uniquely determine the shape of the graph if $p\leq 5$ \rule{0.5em}{0.5em}

{\bf Theorem 4.3} (Theorem 5.3 from \cite{CP})  
{\it Let $T$ be a tree. The eigenvalues of problem (\ref{2.1})--(\ref{2.4}) can be presented as the union of subsequences $\{\lambda_k\}_{k=1}^{\infty}=\mathop{\cup}\limits_{i=1}^{2p-3}\{\lambda_k^{(i)}\}_{k=1}^{\infty}$
with the following asymptotics
\begin{equation}
\label{4.5}
\sqrt{\lambda_k^{(i)}}\mathop{=}\limits_{k\to\infty}\frac{2\pi (k-1)}{l}+\frac{1}{l}\arccos \alpha_{i}+O\left(\frac{1}{k}\right) \ \ {\rm for}  \ \  i=2,3,..., p-1,
\end{equation}
\begin{equation}
\label{4.6}
\sqrt{\lambda_k^{(i+p-2)}}\mathop{=}\limits_{k\to\infty}\frac{2\pi k}{l}
-
\frac{1}{l}\arccos \alpha_{i}+O\left(\frac{1}{k}\right) \ \  {\rm for} \ \  i=2, 3, ..., p-1,
\end{equation}
\begin{equation}
\label{4.7}
\sqrt{\lambda_k^{(1)}}\mathop{=}\limits_{k\to\infty}\frac{\pi (k-1)}{l}+O\left(\frac{1}{k}\right) \ \  
\end{equation}

Here $\alpha_1=1, \alpha_2,\alpha_3, ..., \alpha_{p-1}, \alpha_p=-1$ are the eigenvalues of the discrete \\ Laplacian $\tilde{A}$.}

The proof of the next theorem is mutatis mutandis the proof of Theorem 4.2.

{\bf Theorem 4.4} {\it Let the spectrum  $\{\lambda_k\}_{k=1}^{\infty}$ of problem (\ref{2.1})--(\ref{2.4}) for a simple connected equilateral graph consist of subsequences: $\{\lambda_k\}_{k=1}^{\infty}=\mathop{\cup}\limits_{i=1}^{2p-3}\{\lambda_k^{(i)}\}_{k=1}^{\infty}$
with the asymptotics  
$$
\sqrt{\lambda_k^{(i)}}\mathop{=}\limits_{k\to\infty}\frac{2\pi (k-1)}{l}+\gamma_{i}+O\left(\frac{1}{k}\right) \ \ {\rm for}  \ \  i=2,3,..., p-1,
$$
$$
\sqrt{\lambda_k^{(i+p-2)}}\mathop{=}\limits_{k\to\infty}\frac{2\pi k}{l}
-
\gamma_{i}+O\left(\frac{1}{k}\right) \ \  {\rm for} \ \  i=2, 3, ..., p-1,
$$
$$
\sqrt{\lambda_k^{(1)}}\mathop{=}\limits_{k\to\infty}\frac{\pi (k-1)}{l}+O\left(\frac{1}{k}\right) \ \  
$$
with $0<p\leq 8$. 

Then the graph is a tree and the set $\{\alpha_i\}_{i=2}^{p-1}$ where $\alpha_i=\cos\gamma_i l$ uniquely determines the shape of the tree.}

All possible spectra corresponding to trees of $p\in \{6,7,8\}$ vertices  consist of subsequences (\ref{4.1})--(\ref{4.3}), i.e. $\gamma_i=\frac{1}{l}arccos \ \alpha_i$
where $\{\alpha_i\}_{i=1}^p$ are the eigenvalues of the corresponding discrete Laplacian, i.e. the zeros of one of the polynomials given in \cite{CP}.
The corresponding graphs are also shown in \cite{CP}.

It is clear from the proof of Theorem 4.2  that looking at the first two terms of the asymptotics we can't distinguish two graph only if the numbers of vertices are the same,  the numbers of edges are the same and the sets $\{\alpha_k\}_{k=1}^p$ corresponding to the two graphs coincide. The latter means that the characteristic polynomial $\det(zI-\tilde{A})$ corresponding to one of the  graphs is a multiple of the characteristic polynomial of the other one. All simple connected graphs except of trees with 6 vertices are shown at Figures 5 - 9.

\begin{figure}[H]
 \begin{center}
    \includegraphics[scale= 0.88 ] {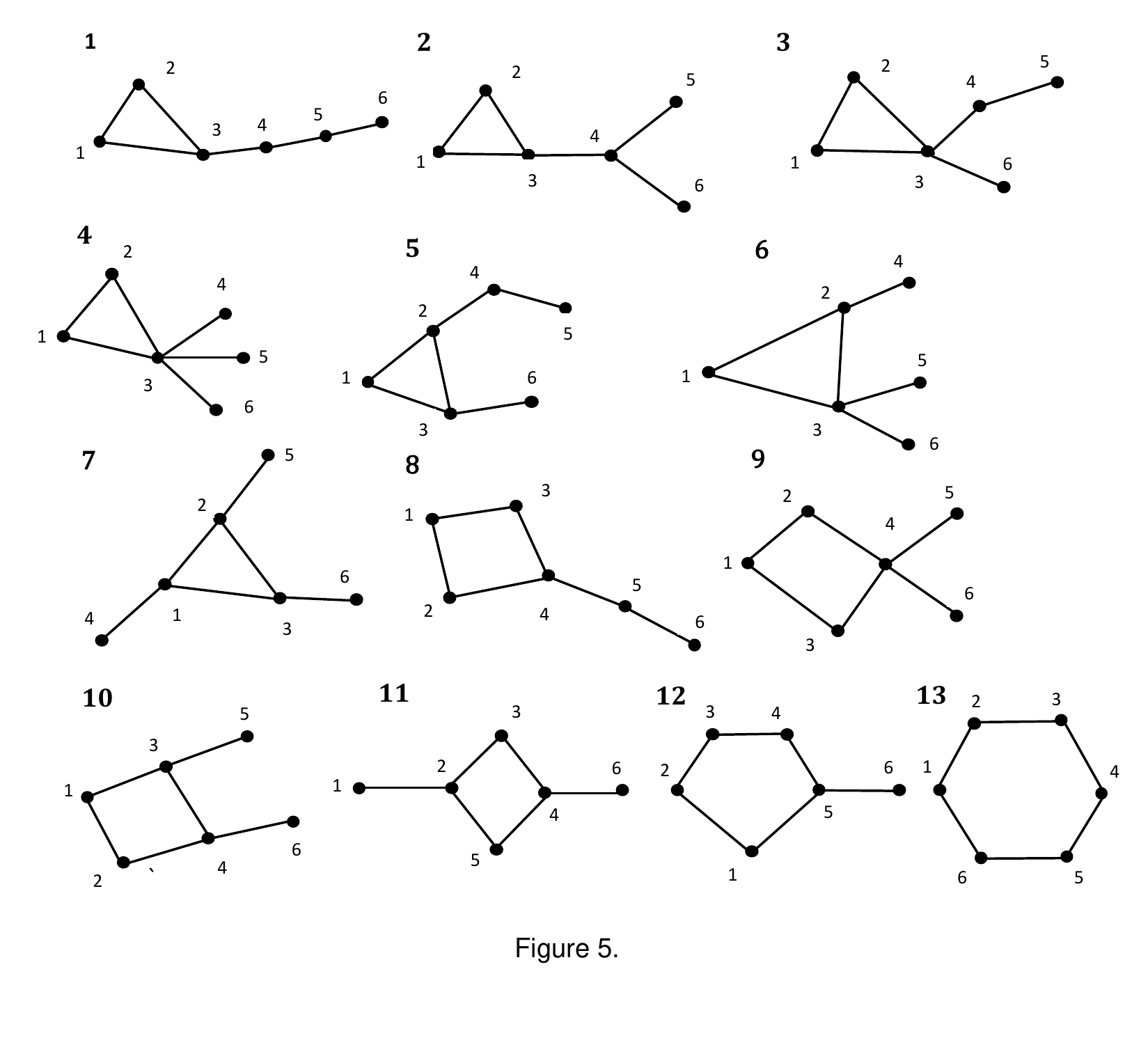}
   \end{center}
\end{figure}

\begin{figure}[H]
 \begin{center}
    \includegraphics[scale= 0.88 ] {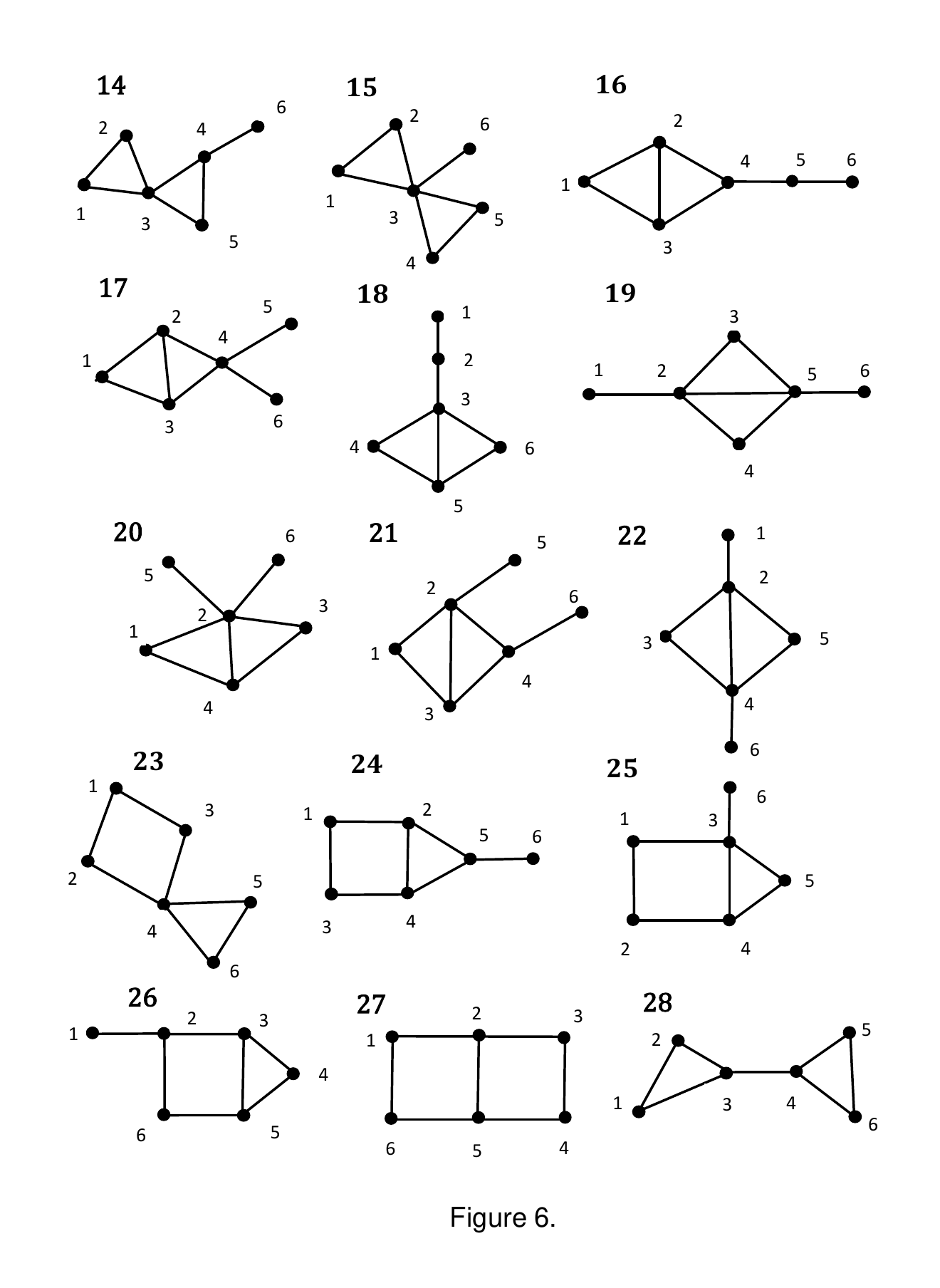}
   \end{center}
\end{figure}

\begin{figure}[H]
 \begin{center}
    \includegraphics[scale= 0.88 ] {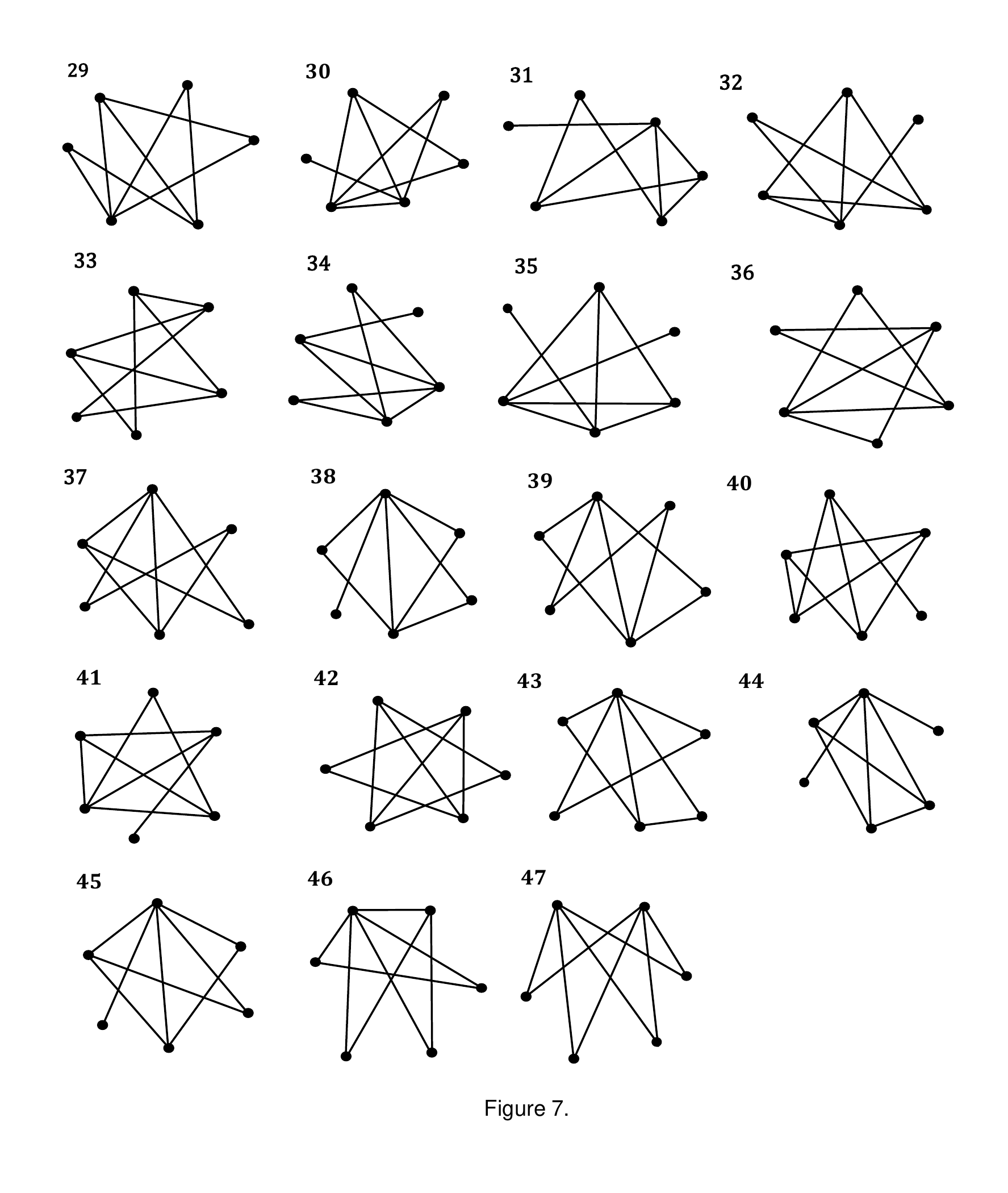}
   \end{center}
\end{figure}

\begin{figure}[H]
 \begin{center}
    \includegraphics[scale= 0.88 ] {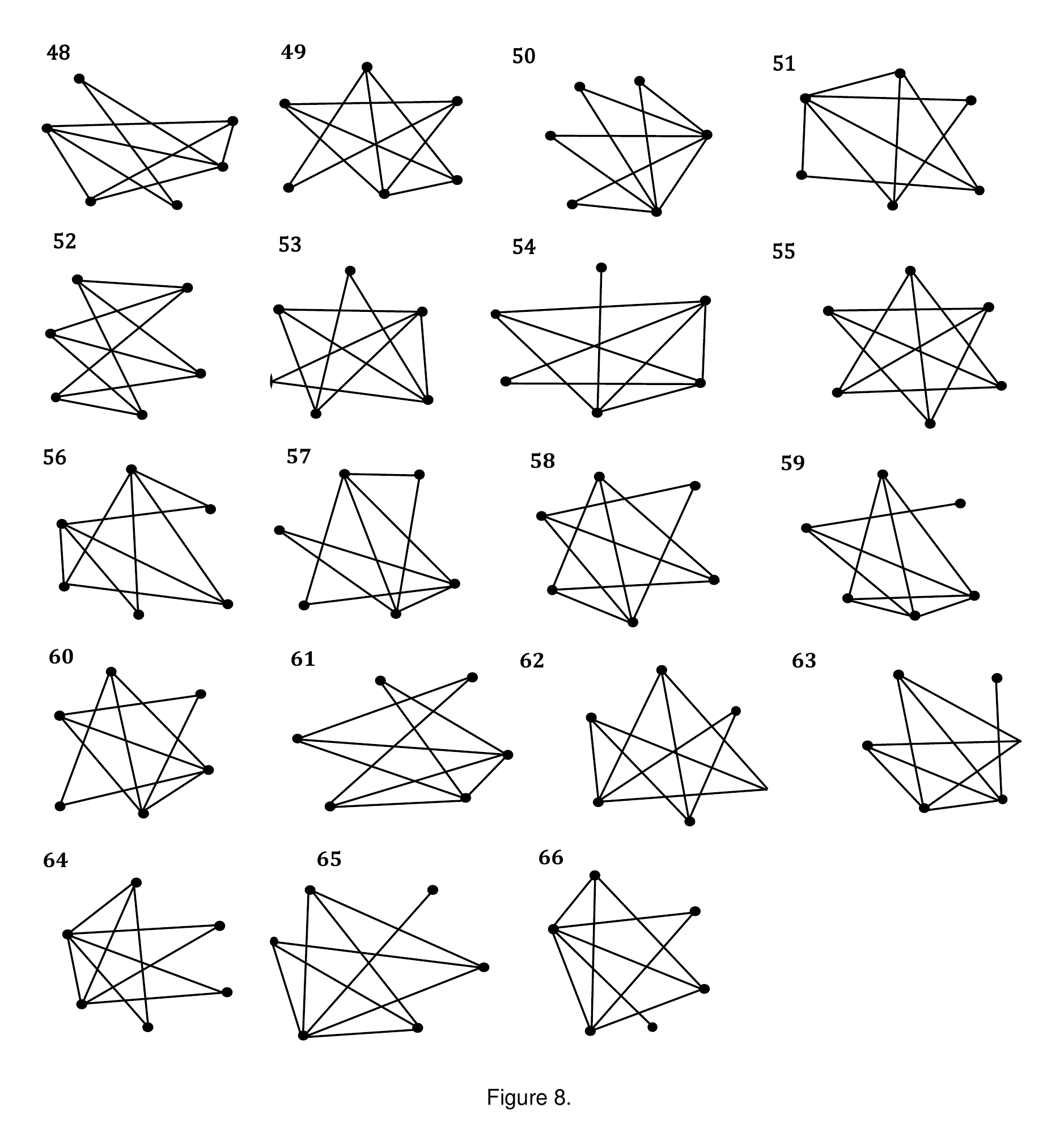}
   \end{center}
\end{figure}

\begin{figure}[H]
 \begin{center}
    \includegraphics[scale= 0.88 ] {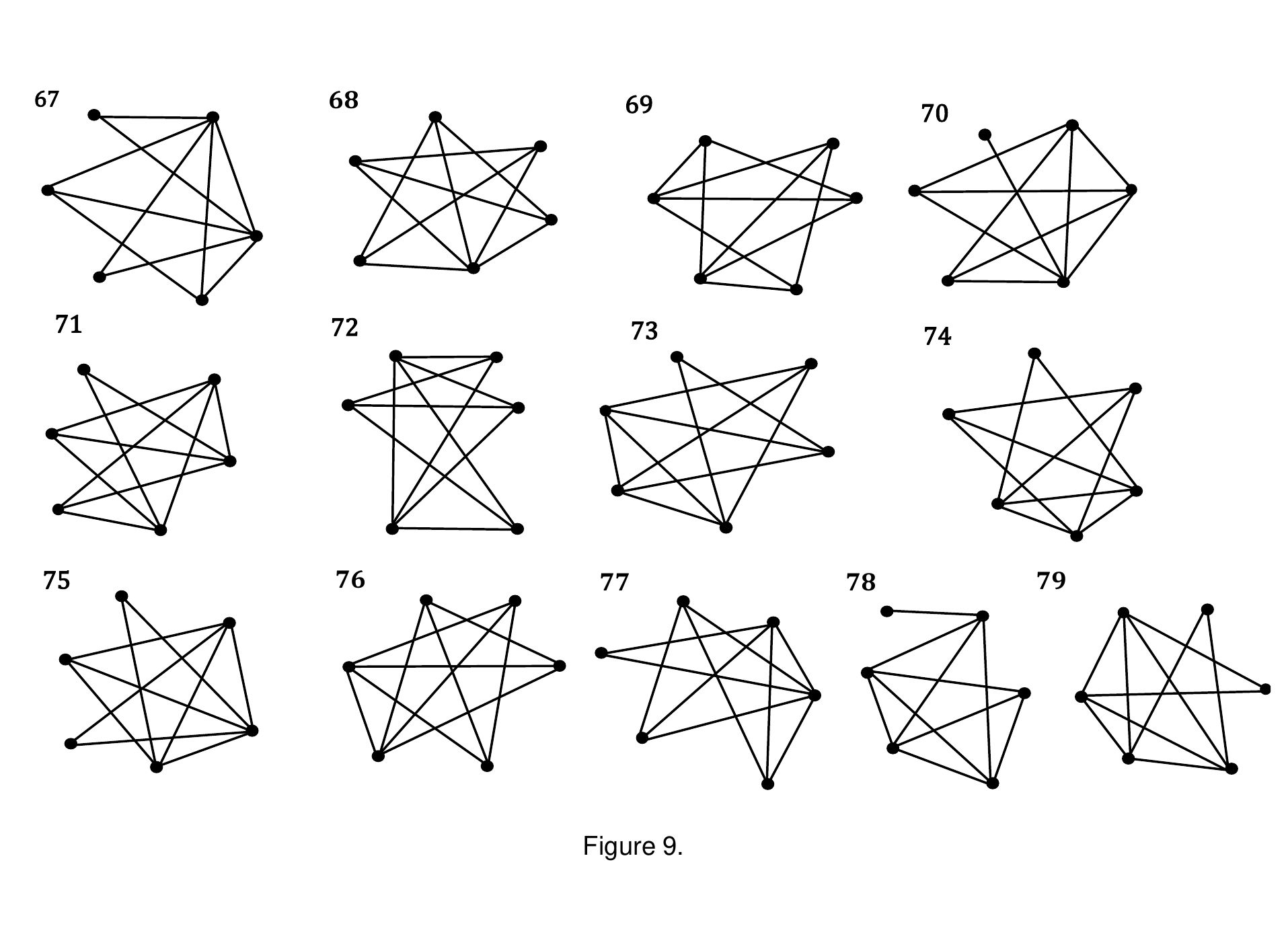}
   \end{center}
\end{figure}

\begin{figure}[H]
 \begin{center}
    \includegraphics[scale= 0.88 ] {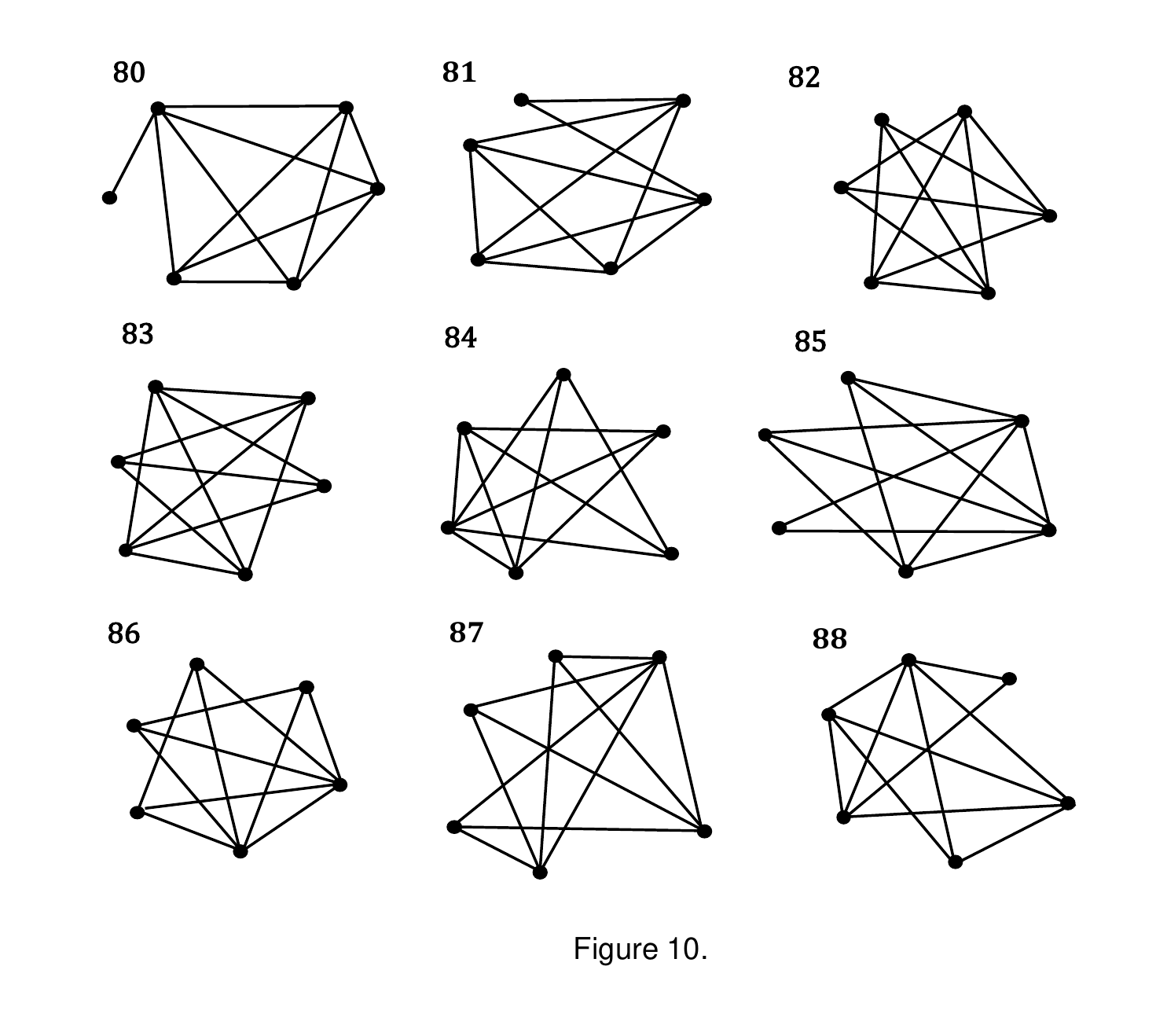}
   \end{center}
\end{figure}

\begin{figure}[H]
 \begin{center}
    \includegraphics[scale= 0.88 ] {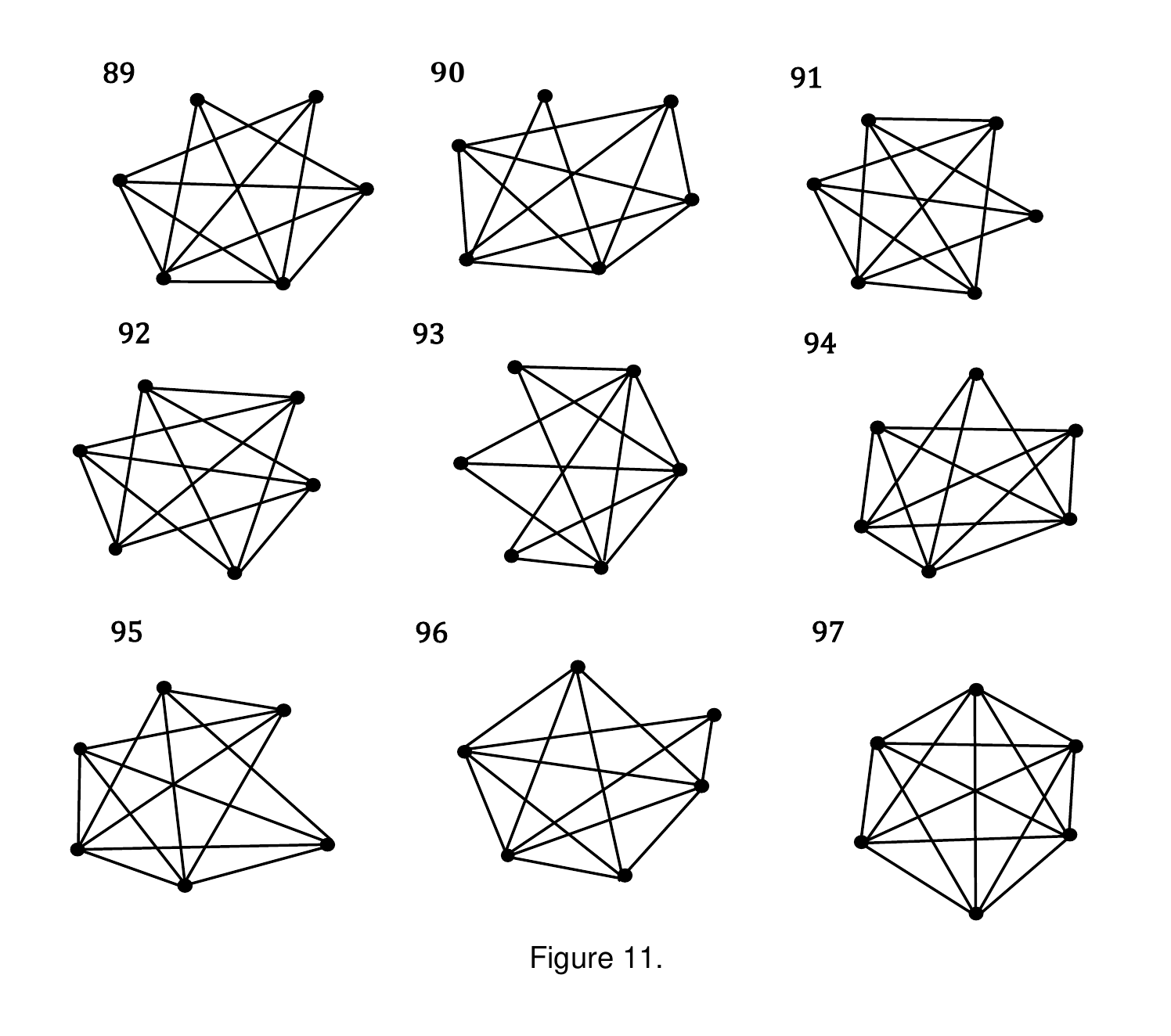}
   \end{center}
\end{figure}

The corresponding determinants of discrete Laplacians are given below  
$$
\phi_1=48z^6-72z^4-8z^3+27z^2+6z-1, \ \ \phi_2=36z^6-49z^4-6z^3+15z^2+4z,
$$
$$
\phi_3=32z^6-44z^4-4z^3+15z^2+2z-1, \ \ \phi_4=20z^6-21z^4-2z^3+3z^2,
$$
$$
\phi_5=36z^6-52z^4-4z^3+19z^2+2z-1, \ \ \phi_6=24z^6-29z^4-2z^3+7z^2,
$$
$$
\phi_7=27z^6-36z^4-2z^3+12z^2-1, \ \ \phi_8=48z^6-72z^4+24z^2, \ \ \phi_9=32z^6-40z^4+8z^2,
$$
$$
\phi_{10}=36z^6-49z^4+14z^2-1, \ \ \phi_{11}=36z^6-49z^4-6z^3+15z^2+4z,
$$
$$
\phi_{11}=36z^6-48z^4+12z^2, \ \ \phi_{12}=48z^6-68z^4-4z^3+21z^2-z-1,
$$
$$
\phi_{13}=64z^6-96z^4-36z^2-4, \ \ 
$$

$$
\phi_{14}=96z^6-116z^4-20z^3+33z^2+8z-1, \ \ \phi_{15}=80z^6-88z^4-16z^3+21z^2+4z-1,
$$
$$
\phi_{16}=108z^6-144z^4-20z^3+44z^2+12z, \ \ \phi_{17}=72z^6-80z^4-12z^3+16z^2+4z,
$$
$$
\phi_{18}=96z^6-124z^4-16z^3+36z^2+8z, \ \ \phi_{19}=64z^6-68z^4-8z^3+12z^2,
$$
$$
\phi_{20}=60z^6-60z^4-8z^3+8z^2, \ \ \phi_{21}=72z^6-83z^4-10z^3+20z^2+2z-1,
$$
$$
\phi_{22}=81z^6-99z^4-12z^3+27z^2+4z-1, \ \ \phi_{23}=128z^6-160z^4-16z^3+40z^2+8z,
$$
$$
\phi_{24}=108z^6-135z^4-8z^3+35z^2, \ \ \phi_{25}=96z^6-112z^4-8z^3+25z^2-1,
$$
$$
\phi_{26}=108z^6-132z^4-12z^3+33z^2+4z-1, \ \ \phi_{27}=
144z^6-184z^4+41z^2-1,
$$
$$
\phi_{28}=144z^6-160z^4-36z^3+43z^2+17z+1,
$$

$$
\phi_{29}=288z^6-308z^4-24z^3+44z^2, \ \ \phi_{30}=192z^6-196z^4-36z^3+35z^2+6z-1,
$$
$$
\phi_{31}=216z^6-228z^4-24z^3+36z^2, \ \ \phi_{32}=216z^6-171z^4-28z^3+7z^2,
$$
$$
\phi_{33}=324z^6-360z^4+36z^2, \ \ \phi_{34}=192z^6-204z^4-32z^3+36z^2+8z,
$$
$$
\phi_{35}=144z^6-145z^4-28z^3+26z^2+4z-1, \ \ \phi_{36}=288z^6-312z^4-48z^3+68z^2+8z-4,
$$
$$
\phi_{37}=288z^6-296z^4-20z^3+53z^2+5z-1, \ \ \phi_{38}=160z^6-148z^4-24z^3+12z^2,
$$
$$
\phi_{39}=256z^6-272z^4-32z^3+48z^2, \ \ \phi_{40}=243z^6-270z^4-36z^3+51z^2+12z,
$$
$$
\phi_{41}=216z^6-237z^4-42z^3+51z^2+12z, \ \ \phi_{42}=324z^6-360z^4-72z^3+84z^2+24z,
$$
$$
\phi_{43}=240z^6-252z^4-56z^3+53z^2+16z, \ \ \phi_{44}=135z^6-126z^4-28z^3+15z^2+4z,
$$
$$
\phi_{45}=180z^6-176z^4-32z^3+27z^2+2z-1, \ \ \phi_{46}=240z^6-252z^4-56z^3+52z^2+16z,
$$
$$
\phi_{47}=256z^6-256z^4,
$$

$$
\phi_{48}=576z^6-580z^4-112z^3+92z^2+24z, \ \ \phi_{49}=648z^6-648z^4-108z^3+104z^2+8z-4,
$$
$$
\phi_{50}=400z^6-336z^4-64z^3, \ \ \phi_{51}=540z^6-516z^4-120z^3+77z^2+20z-1,
$$
$$
\phi_{52}=729z^6-729z^4, \ \ \phi_{53}=576z^6-556z^4-92z^3+69z^2+4z-1,
$$
$$
\phi_{54}=384z^6-360z^4-84z^3+48z^2+12z, \ \ \phi_{55}=729z^6-729z^4-108z^3+108z^2,
$$
$$
\phi_{56}=576z^6-544z^4-64z^3+32z^2, \ \ \phi_{57}=512z^6-480z^4-112z^3+72z^2+12z-4,
$$
$$
\phi_{58}=648z^6-585z^4-120z^3+35z^2+26z, \ \ \phi_{59}=436z^6-435z^4-102z^3+72z^2+30z+3,
$$
$$
\phi_{60}=576z^6-564z^4-144z^3+93z^2+36z+3, \ \ \phi_{61}=576z^6-564z^4-84z^3+60z^2+12z,
$$
$$
\phi_{62}=648z^6-639z^4-72z^3+63z^2, \ \ \phi_{63}=432z^6-411z^4-84z^3+51z^2+12z,
$$
$$
\phi_{64}=480z^6-440z^4-96z^3+48z^2+8z, \ \ \phi_{65}=405z^6-369^4-72z^3+36z^2,
$$
$$
\phi_{66}=360z^6-319z^4-78z^3+32z^2+6z-1, 
$$

$$
\phi_{67}=900z^6-736z^4-200z^3+28z^2+8z, \ \ \phi_{68}=1215z^6-1080^4-270z^3+120z^2+20z-5,
$$
$$
\phi_{69}=1296z^6-1152z^4-288z^3+112z^2+32z, \ \ \phi_{70}=720z^6-597^4-222z^3+15z^2+9z,
$$
$$
\phi_{71}=1152z^6-1008z^4-192z^3+48z^2, \ \ \phi_{72}=1296z^6-1161^4-162z^3+27z^2,
$$
$$
\phi_{73}=1152z^6-1032z^4-228z^3+91z^2+18z-1, \ \ \phi_{74}=1152z^6-1016^4-268z^3+104z^2+28z,
$$
$$
\phi_{75}=960z^6-812z^4-224z^3+63z^2+14z-1, \ \ \phi_{76}=1296z^6-1161^4-252z^3+106z^2+12z-1,
$$
$$
\phi_{77}=1080z^6-933z^4-222z^3+63z^2+12z, \ \ \phi_{78}=768z^6-672z^4-192z^3+63z^2+30z+3,
$$
$$
\phi_{79}=1024z^6-896z^4-252z^3+84z^2+40z+4.
$$

$$
\phi_{80}=1280z^6-992z^4-352z^3+33z^2+28z+3, \ \ \phi_{81}=2048z^6-1664z^4-448z^3+48z^2+16z,
$$
$$
\phi_{82}=2304z^6-1872z^4-528z^3+80z^2+16z, \ \ \phi_{83}=2304z^6-1888z^4-480z^3+53z^2+12z-1,
$$
$$
\phi_{84}=2160z^6-1743z^4-528z^3+83z^2+28z, \ \ \phi_{85}=1800z^6-1392z^4-444z^3+24z^2+12z,
$$
$$
\phi_{86}=2025z^6-1611z^4-576z^3+91z^2+64z+7, \ \ \phi_{87}=2160z^6-1728z^4-432z^3,
$$
$$
\phi_{88}=1920z^6-1528z^4-508z^3+73z^2+40z+3,
$$
$$
\phi_{89}=3600z^6-2649z^4-996z^3+6z^2+36z+3, \ \ \phi_{90}=3200z^6-2328z^4-900z^3-12z^2+36z+4,
$$
$$
\phi_{91}=3840z^6-2864z^4-976z^3+16z^2+16z, \ \ \phi_{92}=4096z^6-3072z^4-1024z^3,
$$
$$
\phi_{93}=3375z^6-2430z^4-864z^3-81z^2, \ \
$$
$$
 \phi_{94}=6000z^6-4095z^4-1746z^3-180z^2+18z+3,\ \ \  \ \phi_{95}=6400z^6-4416z^4-1792z^3-192z^2,
$$
$$
\phi_{96}=1000z^6-6400z^4-3040z^3-528z^2-32z, \ \ \phi_{97}=15625z^6-9375z^4-5000z^3-1125z^2-120z-5,
$$
Comparing these polynomials we obtain that the only pair of such graphs of 6 vertices with the same number of edges and the same set $\{\alpha_i\}_{i=1}^6$      consists of the graphs shown at Fig. 3, i.e. graphs 78 and 79 at Fig. 9. This set $\{\alpha_i\}_{i=1}^6$ is the set of zeros of the polynomial  given by (\ref{4.15}) (see below).
Thus, we arrive at 
 
{\bf Theorem 4.5} {\it 1. Let the spectrum  $\{\lambda_k\}_{k=1}^{\infty}$ of problem (\ref{2.1})--(\ref{2.4}) for a simple connected equilateral graph consist of subsequences:
$\{\lambda_k\}_{k=1}^{\infty}=\mathop{\cup}\limits_{i=1}^{16}\{\lambda_k^{(i)}\}_{k=1}^{\infty}$ which satisfy asymptotics  
\begin{equation}
\label{4.8}
\sqrt{\lambda_k^{(i)}}\mathop{=}\limits_{k\to\infty}\frac{2\pi (k-1)}{l}+\gamma_{i}+O\left(\frac{1}{k}\right) \ \ {\rm for}  \ \  i=1,2,..., 6,
\end{equation}
\begin{equation}
\label{4.9}
\sqrt{\lambda_k^{(i+6)}}\mathop{=}\limits_{k\to\infty}\frac{2\pi k}{l}
-
\gamma_{i}+O\left(\frac{1}{k}\right) \ \  {\rm for} \ \  i=1, 2, ..., 6,
\end{equation}
\begin{equation}
\label{4.10}
\sqrt{\lambda_k^{(12+i)}}\mathop{=}\limits_{k\to\infty}\frac{\pi k}{l}+O\left(\frac{1}{k}\right) \ \  {\rm for} \ \ i=1, 2, 3, 4.  \ \
\end{equation}
where the set $\{\alpha_i\}_{i=1}^6=\{\cos\gamma_i l\}_{i=1}^6$ does not coincide with the set of zeros  of the polynomial
\begin{equation}
\label{4.15}
\phi=256z^6-224z^4-64z^3+21z^2+10z+1=(z-1)(4z+1)^3(4z^2+z-1).
\end{equation}  
Then the set $\{\alpha_i\}_{i=1}^6$ uniquely determines the shape of the graph.

2. The two nonisomorphic graphs shown on Fig 3 have the same first and second terms of  
 asymptotics (\ref{4.8})--(\ref{4.10}) and the characteristic polynomial (\ref{4.15}).

}

It should be mentioned that in \cite{P} a pair of cospectral trees of 9 vertices is shown.  

\section{Spectral problems with the Dirichlet boundary conditions at the pendant vertices}
\setcounter{equation}{0} \hskip0.25in

In this section we consider spectral problem (\ref{2.1})--(\ref{2.3}), (\ref{2.5})  ($r=p_{pen}$), i.e. we impose the Dirichlet conditions at all the pendant vertices. We call {\it leaves} the edges of the graph incident with its pendant vertices  and {\it interior subgraph} the subgraph obtained by deleting all the pendant vertices and all the leaves. It is clear that in general we cannot expect uniqueness of the graph corresponding to the given spectrum of problem (\ref{2.1})--(\ref{2.4}), (\ref{2.5}). In the cases where such graph is not unique we can judge about the shape of the interior subgraph. Below we consider this spectral problem for some concrete graphs.

{\bf 1. Double star graph.}

We consider the graph shown at Fig. 12. 

{\bf Theorem 5.1} {\it Let the degrees $m$ and $n$ of the  interior vertices of the double star graph of Fig. 12  satisfy the inequalities $m\geq 1$, $n\geq 1$ and $mn>1$. Then the  
spectrum of problem (\ref{2.1})--(\ref{2.3}), (\ref{2.5}) for this  double star graph consists of the subsequences with the asymptotics 
$$
\sqrt{\lambda_k^{(1)}}\mathop{=}\limits_{k\to\infty}\frac{2\pi (k-1)}{l}+\frac{1}{l}\arccos \frac{1}{\sqrt{mn}} +O\left(\frac{1}{k}\right) \ \ {\rm for}  \ \  \ \ k\in \mathbb{N},
$$
$$
\sqrt{\lambda_k^{(2)}}\mathop{=}\limits_{k\to\infty}\frac{2\pi k}{l}-\frac{1}{l}\arccos \frac{1}{\sqrt{mn}} +O\left(\frac{1}{k}\right) \ \ {\rm for}  \ \  \ \ k\in \mathbb{N},
$$
$$
\sqrt{\lambda_k^{(3)}}\mathop{=}\limits_{k\to\infty}\frac{(2k-1)\pi}{l}+\frac{1}{l}\arccos \frac{1}{\sqrt{mn}} +O\left(\frac{1}{k}\right) \ \ {\rm for}  \ \  \ \ k\in \mathbb{N},
$$
$$
\sqrt{\lambda_k^{(4)}}\mathop{=}\limits_{k\to\infty}\frac{(2k-1)\pi}{l}-\frac{1}{l}\arccos \frac{1}{\sqrt{mn}} +O\left(\frac{1}{k}\right) \ \ {\rm for}  \ \   \ \ k\in \mathbb{N},
$$
$$
\sqrt{\lambda_k^{(i)}}\mathop{=}\limits_{k\to\infty}\frac{\pi k}{l}+O\left(\frac{1}{k}\right)  \ \ {\rm for} \ \  i=5,6,..., m+n+1 \ \ {\rm and} \ \ k\in\mathbb{N}.
$$
}

\begin{figure}[H]
  \begin{center}
    \includegraphics[scale= 0.9] {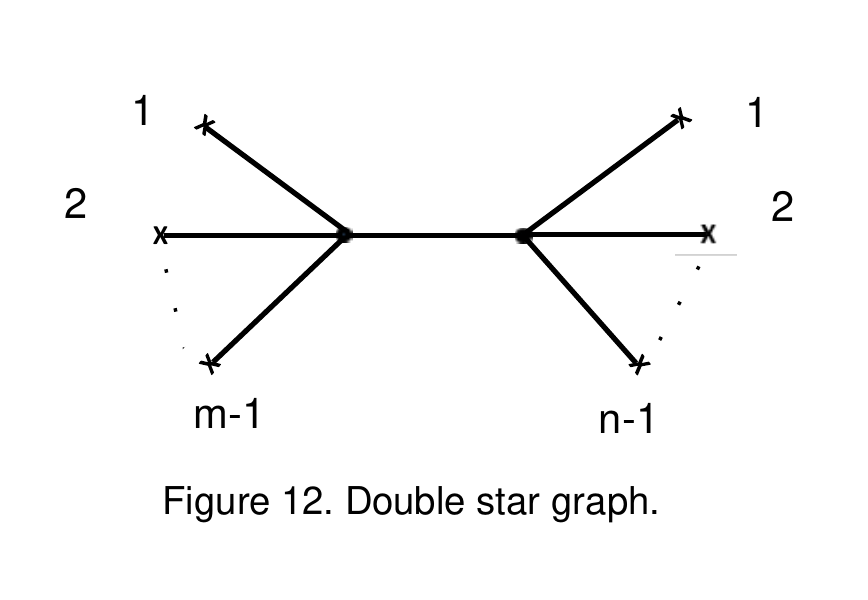}
   \end{center}
\end{figure}

{\bf Proof} The interior subgraph in this case is $P_2$, 
$
\hat{D}_G=diag\{m, n\},
$
$$
\hat{A}=\left(\matrix{0 & 1 \cr 1 & 0 }\right), 
$$ 
and $P_{G,\hat{G}}=1-mnz^2$. Thus, $\alpha_1=-\frac{1}{\sqrt{mn}}$ and $\alpha_2=\frac{1}{\sqrt{mn}}$. Applying Theorem 3.1 we finish the proof \rule{0.5em}{0.5em}

{\bf Theorem 5.2} {\it Let the spectrum of problem (\ref{2.1})--(\ref{2.3}), (\ref{2.5}) consist of  subsequences with the asymptotics 
$$
\sqrt{\lambda_k^{(1)}}\mathop{=}\limits_{k\to\infty}\frac{2\pi (k-1)}{l}+\gamma +O\left(\frac{1}{k}\right) \ \ {\rm for}  \ \  \ \ k\in \mathbb{N},
$$
$$
\sqrt{\lambda_k^{(2)}}\mathop{=}\limits_{k\to\infty}\frac{2\pi k}{l}-\gamma +O\left(\frac{1}{k}\right) \ \ {\rm for}  \ \  \ \ k\in \mathbb{N},
$$
$$
\sqrt{\lambda_k^{(3)}}\mathop{=}\limits_{k\to\infty}\frac{(2k-1)\pi}{l}+\gamma +O\left(\frac{1}{k}\right) \ \ {\rm for}  \ \  \ \ k\in \mathbb{N},
$$
$$
\sqrt{\lambda_k^{(4)}}\mathop{=}\limits_{k\to\infty}\frac{(2k-1)\pi}{l}-\gamma +O\left(\frac{1}{k}\right) \ \ {\rm for}  \ \   \ \ k\in \mathbb{N},
$$
$$
\sqrt{\lambda_k^{(i)}}\mathop{=}\limits_{k\to\infty}\frac{\pi k}{l}+O\left(\frac{1}{k}\right)  \ \ {\rm for} \ \  i=5,6,..., s \ \ {\rm and} \ \ k\in\mathbb{N}.
$$

 Then these asymptotics uniquely determine the shape of the graph as a double star with the number of leaves incident with the interior vertices   $m-1$ and $n-1$ (see Fig. 12) where the pair of natural numbers $m$ and $n$ compose a solution of the system of equations 
 \begin{equation}
 \label{5.11}
  m+n=s-1, \ \ mn=(\cos \gamma l)^{-2}.
 \end{equation}
 This system possesses two solutions which correspond to isomorphic graphs.  }

{\bf Proof} 
By Theorem 5.4 in \cite{CaP} we conclude that $|\lambda^{(j)}_k-{\tilde \lambda}^{(j)}_k|\leq C<\infty$ where  ${\tilde \lambda}^{(j)}_k$ are the eigenvalues of problem (\ref{2.1})--(\ref{2.3}), (\ref{2.5})  on the same graph with $q_j\equiv 0$ for all $j$ and therefore, presence of the potentials does not influence the first and the second terms of the asymptotics. By conditions of our theorem $\{\lambda_k\}_{k=1}^{\infty}=\mathop{\cup}\limits_{i=1}^{m+n+1}\{\lambda_k^{(i)}\}_{k=1}^{\infty}$ is the spectrum of problem (\ref{2.1})--(\ref{2.3}), (\ref{2.5}) on a simple connected graph, therefore the spectrum of the same problem with 0 potentials can be given as the union of subsequences such that

$$
\sqrt{\tilde{\lambda}_k^{(1)}}=\frac{2\pi (k-1)}{l}+\frac{1}{l}\arccos \frac{1}{\sqrt{mn}}  \ \ {\rm for}  \ \  \ \ k\in \mathbb{N},
$$
$$
\sqrt{\tilde{\lambda}_k^{(2)}}=\frac{2\pi k}{l}-\frac{1}{l}\arccos \frac{1}{\sqrt{mn}}  \ \ {\rm for}  \ \  \ \ k\in \mathbb{N},
$$
$$
\sqrt{\tilde{\lambda}_k^{(3)}}=\frac{(2k-1)\pi}{l}+\frac{1}{l}\arccos \frac{1}{\sqrt{mn}}  \ \ {\rm for}  \ \  \ \ k\in \mathbb{N},
$$
$$
\sqrt{\tilde{\lambda}_k^{(4)}}=\frac{(2k-1)\pi}{l}-\frac{1}{l}\arccos \frac{1}{\sqrt{mn}}  \ \ {\rm for}  \ \   \ \ k\in \mathbb{N},
$$
$$
\sqrt{\tilde{\lambda}_k^{(i)}}=\frac{\pi k}{l}  \ \ {\rm for} \ \  i=5,6,..., m+n+1 \ \ {\rm and} \ \ k\in\mathbb{N}.
$$
The set  $\mathop{\cup}\limits_{i=1}^{m+n+1}\{\tilde{\lambda}_k^{(i)}\}_{k=1}^{\infty}$ is the spectrum of problem (\ref{2.1})--(\ref{2.3}), (\ref{2.5}) on the double star graph of Fig. 12 with zero potentials on the edges. The two solutions of system (\ref{5.11}) correspond to the isomorphic graphs where $n$ and $m$ are swapped \rule{0.5em}{0.5em}

Below we show that if the graph is more complicated the asymptotics of problem (\ref{2.1})--(\ref{2.3}), (\ref{2.5}) do not determine the shape of the graph uniquely.

{\bf 2. Caterpillar graph.}  

Here we consider the graph of Fig. 13. 

{\bf Theorem 5.3} {\it The spectrum of problem (\ref{2.1})--(\ref{2.3}), (\ref{2.5}) for the caterpillar graph of Fig. 13 consists of subsequences which behave asymptotically as follows
\begin{equation}
\label{5.12}
\sqrt{\lambda_k^{(1)}}\mathop{=}\limits_{k\to\infty}\frac{2\pi (k-1)}{l}+\gamma+O\left(\frac{1}{k}\right) \ \ {\rm for}  \ \  \ \ k\in \mathbb{N},
\end{equation}
\begin{equation}
\label{5.13}
\sqrt{\lambda_k^{(2)}}\mathop{=}\limits_{k\to\infty}\frac{2\pi k}{l}-\gamma +O\left(\frac{1}{k}\right) \ \ {\rm for}  \ \  \ \ k\in \mathbb{N},
\end{equation}
\begin{equation}
\label{5.14}
\sqrt{\lambda_k^{(3)}}\mathop{=}\limits_{k\to\infty}\frac{(2k-1)\pi}{l}+ \gamma +O\left(\frac{1}{k}\right) \ \ {\rm for}  \ \  \ \ k\in \mathbb{N},
\end{equation}
\begin{equation}
\label{5.15}
\sqrt{\lambda_k^{(4)}}\mathop{=}\limits_{k\to\infty}\frac{(2k-1)\pi}{l}-\gamma +O\left(\frac{1}{k}\right) \ \ {\rm for}  \ \   \ \ k\in \mathbb{N},
\end{equation}
\begin{equation}
\label{5.16}
\sqrt{\lambda_k^{(5)}}\mathop{=}\limits_{k\to\infty}\frac{(k-1/2)\pi}{l} +O\left(\frac{1}{k}\right) \ \ {\rm for}  \ \   \ \ k\in \mathbb{N},
\end{equation}
\begin{equation}
\label{5.17}
\sqrt{\lambda_k^{(i)}}\mathop{=}\limits_{k\to\infty}\frac{\pi k}{l}+O\left(\frac{1}{k}\right)  \ \ i=6,7,..., s
\end{equation}
where 
$$
\gamma :=\frac{1}{l}\arccos \sqrt{\frac{m_1+m_3}{m_1m_2m_3}}, \ \ s=m_1+m_2+m_3
$$
and $m_i$ ($i=1,2,3$) are the degrees of the interior vertices $v_1$, $v_2$, $v_3$, respectively. }

\begin{figure}[H]
  \begin{center}
    \includegraphics[scale= 0.9] {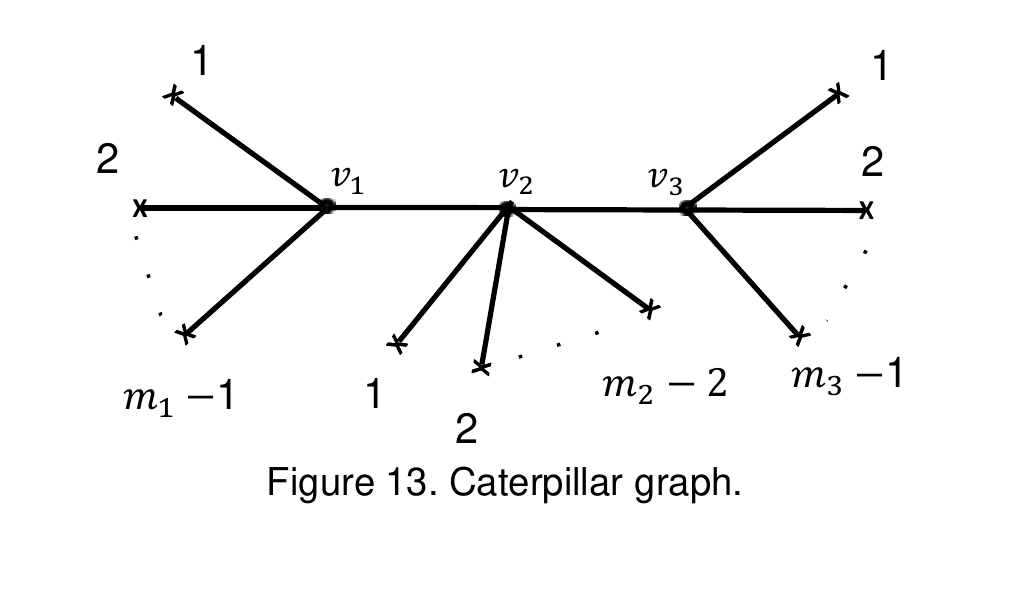}
   \end{center}
\end{figure}

{\bf Proof} 
The interior subgraph in this case is $P_3$, 
$$
\hat{D}_G=diag\{m_1, m_2, m_3\},
$$
$$
\hat{A}=\left(\matrix{0 & 1 & 0 \cr 1 & 0 & 1 \cr 0 & 1 & 0}\right), 
$$
and $P_{G,\hat{G}}=(m_1+m_3)z-m_1m_2m_3z^3$. Thus, the zeros of this polynomial are $\alpha_1=-\sqrt{\frac{m_1+m_3}{m_1m_2m_3}}$,  $\alpha_2=0$ and $\alpha_3=\sqrt{\frac{m_1+m_3}{m_1m_2m_3}}$. Applying Theorem 3.1 we finish the proof \rule{0.5em}{0.5em}

{\bf Theorem 5.4}  {\it Let the spectrum of problem (\ref{2.1})--(\ref{2.3}), (\ref{2.5}) consist of subsequences with the asymptotics 
(\ref{5.12})--(\ref{5.17}).

Then the graph is the caterpillar graph of Fig. 13 with the interior subgraph $P_3$ and the numbers of leaves incident with the interior vertices are $m_1-1$, $m_2-2$, $m_3-1$ where the triplet $m_1\geq  1$, $m_2\geq 2$,  $m_3\geq 1$ is a solution of system 
\begin{equation}
\label{ 5.26}
m_1+m_2+m_3=s,
\end{equation}
\begin{equation}
\label{5.27}
\frac{m_1+m_3}{m_1m_2m_3}=\cos^2\gamma l.
\end{equation}
}

{\bf Proof}  The proof is mutatis mutandis the proof of Theorem 5.2 \rule{0.5em}{0.5em}

{\bf Remark} 1. In some cases the graph corresponding to the given $\gamma$ obtained from the asimptotics and to the given number of edges $(m_1+m_2+m_3-2)$ is unique. For example, if the number of edges is $m_1+m_2+m_3-2=5$ and $\cos^2\gamma l=\frac{5}{12}$ then 
the graph is of the form  
shown at Fig. 14.

\begin{figure}[H]
  \begin{center}
    \includegraphics[scale= 0.9] {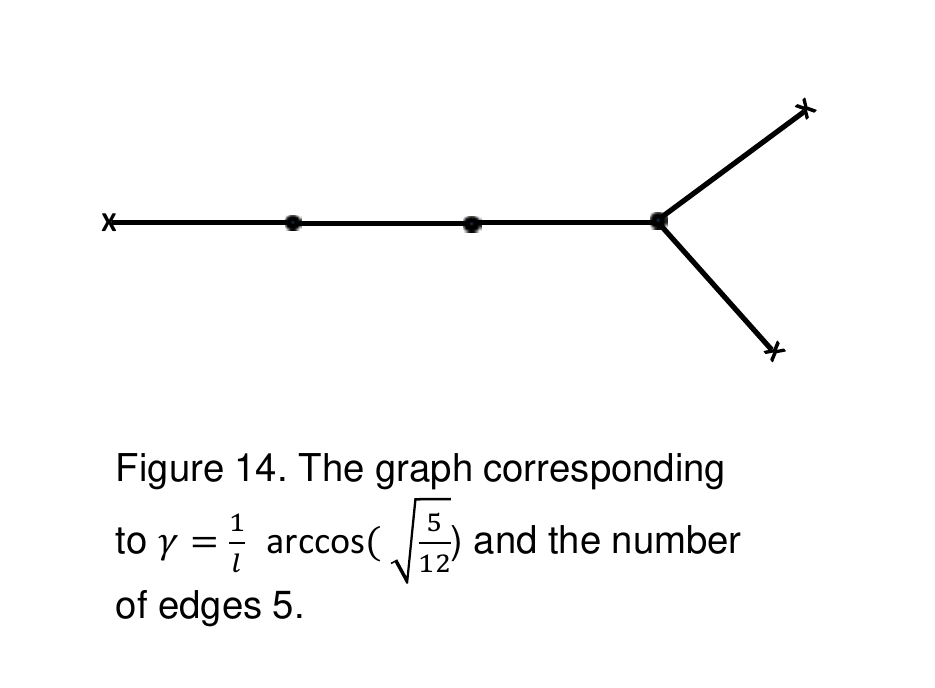}
   \end{center}
\end{figure}

2. If the number of edges is $m_1+m_2+m_3-2=18$ and $\cos^2\gamma l=\frac{1}{16}$ then we have two graphs of Fig. 15 corresponding to $m_1=m_3=8$, $m_2=4$ and $m_1=m_3=2$, $m_2=16$, respectively.

\begin{figure}[H]
  \begin{center}
    \includegraphics[scale= 0.88 ] {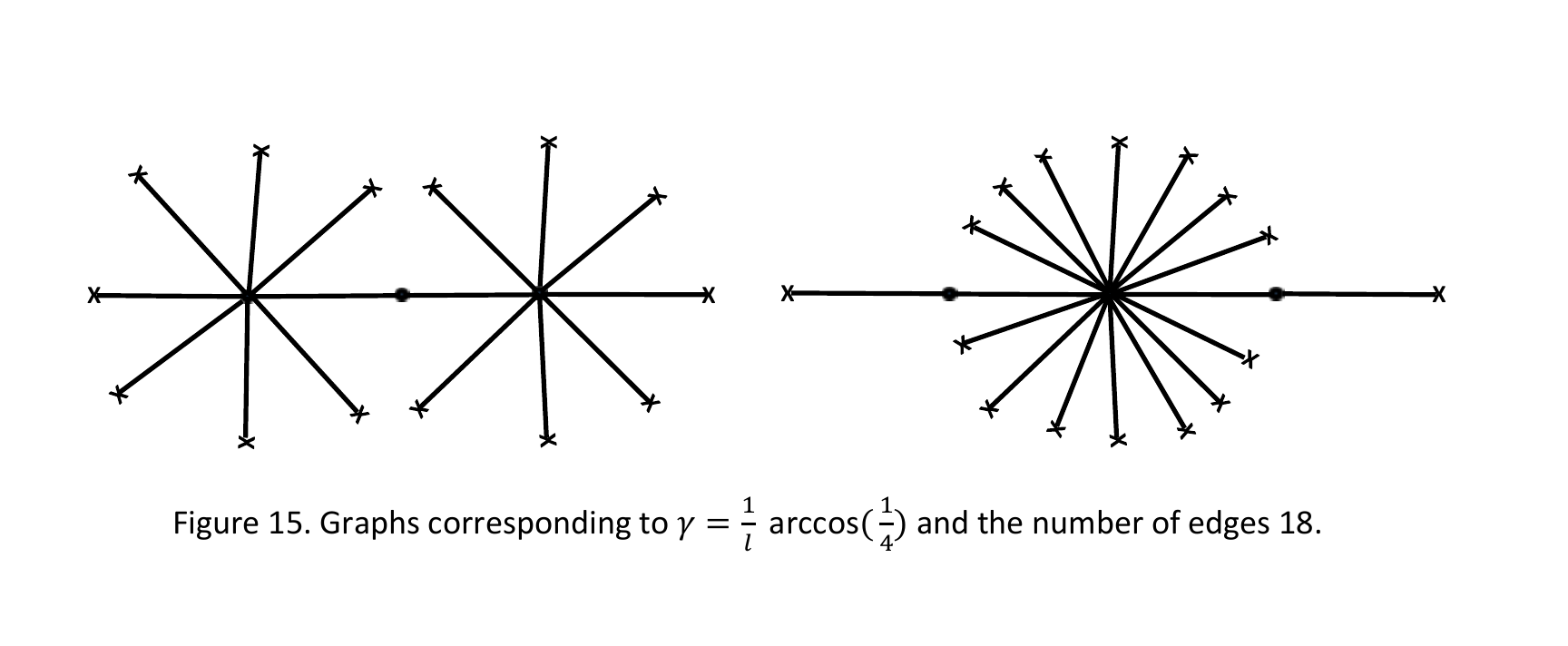}
   \end{center}
\end{figure}

{\bf 3. Graph of a decorated triangle}

{\bf Theorem 5.5}   
{\it The spectrum of  problem (\ref{2.1})--(\ref{2.3}), (\ref{2.5}) for the graph of Fig. 16 consists of subsequences which behave asymptotically as follows:
\begin{equation}
\label{5.28}
\sqrt{\lambda_k^{(i)}}\mathop{=}\limits_{k\to\infty}\frac{2\pi (k-1)}{l}+\gamma_i+O\left(\frac{1}{k}\right) \ \ {\rm for}  \ \  \ \ k\in \mathbb{N} \ \ i=1,2,3
\end{equation}
\begin{equation}
\label{5.29}
\sqrt{\lambda_k^{(i)}}\mathop{=}\limits_{k\to\infty}\frac{2\pi k}{l}-\gamma_i+O\left(\frac{1}{k}\right) \ \ {\rm for}  \ \  \ \ k\in \mathbb{N} \ \ i=4,5,6
\end{equation}
\begin{equation}
\label{5.30}
\sqrt{\lambda_k^{(i)}}\mathop{=}\limits_{k\to\infty}\frac{2\pi k}{l}+O\left(\frac{1}{k}\right) \ \ {\rm for}  \ \  \ \ k\in \mathbb{N} \ \ i=7,8,..., s,
\end{equation}
where $s=m_1+m_2+m_3$, $m_i$s are the degrees of the interior vertices,  $\gamma_i:=\frac{1}{l}\arccos \tau_i$ ($i=1, 2, 3$) and $\tau_i$  are the zeros of the polynomial
$$
\phi=m_1m_2m_3z^3-sz-2.
$$
}

\begin{figure}[H]
  \begin{center}
    \includegraphics[scale= 0.88 ] {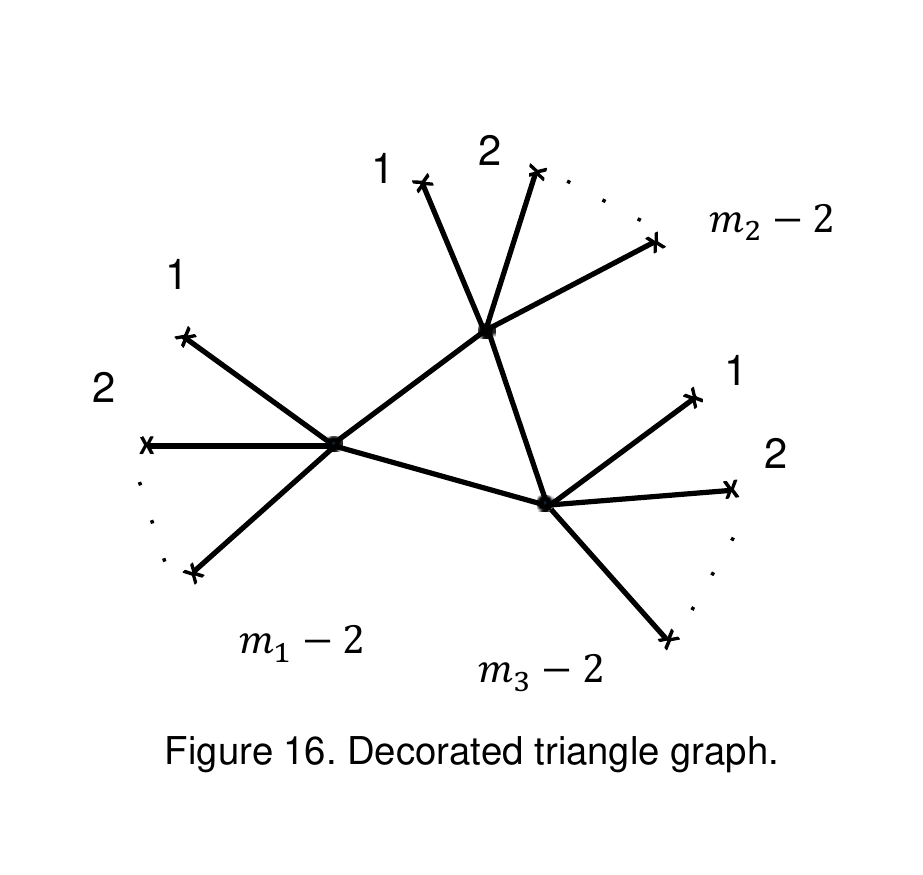}
   \end{center}
\end{figure}

{\bf Proof} The interior subgraph in this case is $C_3$, 
 the degree matrix is
$\hat{D}_G=diag\{m_1, m_2, m_3\},
$
$$
\hat{A}=\left(\matrix{0 & 1 & 1 \cr 1 & 0 & 1 \cr 1 & 1 & 0}\right), 
$$ 
and $P_{G,\hat{G}}=2+(m_1+m_2+m_3)z-m_1m_2m_3z^3$.  Applying Theorem 3.1 we finish the proof.

{\bf Theorem 5.6} {\it Let the spectrum of problem (\ref{2.1})--(\ref{2.3}), (\ref{2.5}) consists of subsequences with the asymptotics 
described by (\ref{5.28})--(\ref{5.30}).

Then the interior subgraph is $C_3$. The numbers $m_i-2$ ($i=1,2,3$) where ($m_1$, $m_2$, $m_3$) is a solution of the system of equations
\begin{equation}
\label{5.32}
m_1m_2m_3=\frac{2}{\tau_1\tau_2\tau_3}
\end{equation}
\begin{equation}
\label{5.33}
m_1+m_2+m_3=s,
\end{equation}
where $m_i$s are the degrees of the interior vertices and $\gamma_i=\frac{1}{l}\arccos \tau_i$. }

{\bf Proof} The proof of this theorem is mutatis mutandis the proof of Theorem 5.2 \rule{0.5em}{0.5em}

{\bf Remark} The solution of the system (\ref{5.32}), (\ref{5.33}) is unique in case of $s=9$ and $\tau_1\tau_2\tau_3=\frac{2}{27}$. In this case  asymptotics (\ref{5.28}) -- (\ref{5.30}) uniquely determine the shape of the graph. Each vertex of the interior subgraph has one incident leaf. In case of $s=10$ and $\tau_1\tau_2\tau_3=\frac{1}{18}$  the solution of system (\ref{5.32}), (\ref{5.33})
is not unique but all the solutions lead to isomorphic graphs. Thus in this case also the asymptotics uniquely determine the shape of the graph. However, in case of   $s=14$ and $\tau_1\tau_2\tau_3=\frac{1}{36}$ the system (\ref{5.32}), (\ref{5.33}) has solutions $m_1=m_2=3$, $m_3=8$ and 
$m_1=m_2=6$, $m_3=2$ which correspond to nonisomorphic graphs shown at Fig 17. 

\begin{figure}[H]
  \begin{center}
    \includegraphics[scale= 0.88 ] {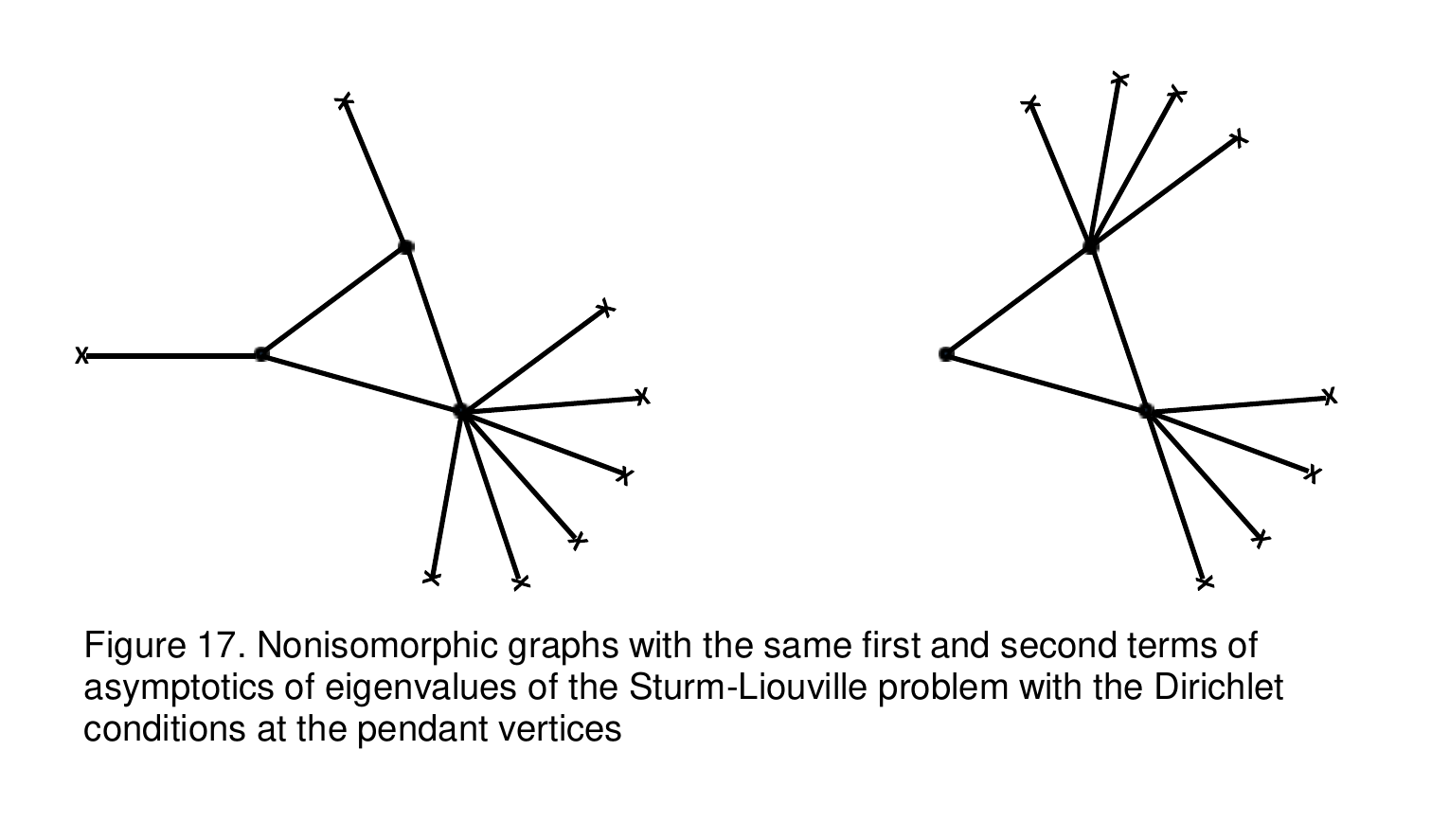}
   \end{center}
\end{figure}


\noindent\textbf{Acknowledgements}\\
\textit{The author is grateful to the Ministry of Education and Science of Ukraine for the support in completing the work “Artificial porous materials as a basis for creating the novel biosensors” \\\ (https://mon.gov.ua/storage/app/uploads/public/61f/944/3d1/61f9443d16f00443598040.pdf).}

.


\end{document}